\documentclass[aps, amsmath,amssymb,prb,twocolumn,superscriptaddress]{revtex4-2}

\usepackage{graphicx}
\usepackage{bm}
\usepackage{hyperref}
\usepackage{dsfont}
\usepackage{color}
\usepackage{xspace}
\usepackage[normalem]{ulem}
\usepackage[compat=1.0.0]{tikz-feynman}
\renewcommand{\i}{\bm{i}}
\renewcommand{\j}{\bm{j}}
\def\Tr{\mathop{\mathrm{Tr}}}
\def\Pf{\mathop{\mathrm{Pf}}}

\newcommand{\eg}[0]{e.g.\@\xspace}
\newcommand{\ie}[0]{i.e.\@\xspace}

\usepackage{ifthen}
\newboolean{draft}

\setboolean{draft}{true}   
\ifthenelse{\boolean{draft}}{%
}{%
}%
\ifthenelse{\boolean{draft}}{%
}{%
}%
\ifthenelse{\boolean{draft}}{%
}{%
}%
\ifthenelse{\boolean{draft}}{%
}{%
}%
\ifthenelse{\boolean{draft}}{%
  \newcommand{\rem}[1]{{\textcolor{red}{\sout{#1}}}}%
}{%
  \newcommand{\rem}[1]{}%
}%

\begin{document}

\title{Valence-bond solid to antiferromagnet transition in the two-dimensional Su-Schrieffer-Heeger model by Langevin dynamics}

\author{A. G\"otz}
\affiliation{\mbox{Institut f\"ur Theoretische Physik und Astrophysik, Universit\"at W\"urzburg, 97074 W\"urzburg, Germany}}

\author{S. Beyl}
\affiliation{\mbox{Institut f\"ur Theoretische Physik und Astrophysik, Universit\"at W\"urzburg, 97074 W\"urzburg, Germany}}

\author{M. Hohenadler}
\affiliation{\mbox{Institut f\"ur Theoretische Physik und Astrophysik, Universit\"at W\"urzburg, 97074 W\"urzburg, Germany}}

\author{F. F. Assaad}
\affiliation{\mbox{Institut f\"ur Theoretische Physik und Astrophysik, Universit\"at W\"urzburg, 97074 W\"urzburg, Germany}}
\affiliation{\mbox{W\"urzburg-Dresden Cluster of Excellence ct.qmat, Am Hubland, 97074 W\"urzburg, Germany}}

\begin{abstract}
    The two-dimensional Su-Schrieffer-Heeger model of electrons coupled to
    quantum phonons is investigated using Langevin dynamics within the framework
    of auxiliary-field quantum Monte Carlo. Based on an explicit determination
    of the density of zeros of the fermion determinant, it is argued that the
    method is efficient in the challenging adiabatic limit. Large-scale
    simulations at the O(4)-symmetric point establish that the ground state of
    the 2D SSH model undergoes a transition from a $(\pi,\pi)$ valence bond
    solid to an antiferromagnet with increasing phonon frequency, yet still in
    the adiabatic regime.  The single-particle spectrum illustrates the
    renormalization of the electronic band and suggests the existence of a
    gapped polaronic band, whereas the particle-hole channels show gapless modes
    associated with long-range bond and magnetic order, respectively. The
    simulations are supplemented with a mean-field analysis and a
    self-consistent Born approximation.
\end{abstract}

\maketitle

\section{Introduction}

The coupling of electrons or spins to phonons can generate many fascinating
states of matter. Apart from superconductivity \cite{Bardeen57}, this also includes phases that break
lattice symmetries such as
charge-density wave states (CDW)  and various flavors of valence-bond solid
(VBS) states \cite{Hohenadler18_rev}.  In spin systems, the coupling to phonons
can generate quantum phase transitions between
antiferromagnetic (AFM)  and VBS states \cite{Weber21}.  A particularly
intriguing aspect is the possibility of realizing quantum phase transitions
beyond the Landau-Ginzburg-Wilson paradigm---connecting two states with
different local order parameters---in models relevant for materials.

The Debye frequency $\omega_\text{D}$ is typically much  smaller than the
Fermi  energy  $\epsilon_\text{F}$.  This separation of energy scales underlies
Migdal's  theorem \cite{Migdal58,Bardeen57}, which provides a small
parameter,  $\hbar \omega_\text{D}/\epsilon_\text{F}$, to  justify  perturbative approaches
to the electron-phonon problem. Quantum Monte Carlo (QMC) simulations offer the
possibility to take a step beyond  perturbative approaches and thereby
investigate competing instabilities  \cite{Esterlis18}.  In fact, the generic
electron-phonon problem  does not suffer from a negative
sign problem, irrespective of lattice geometry and band filling.
In particular, for each space-time configuration of phonon fields,
time-reversal symmetry ensures that the eigenvalues of the fermion
determinant come in complex conjugate pairs  \cite{Wu04}. As such, it
should be a technically simple problem. However, this is not the case.
First, the argument for the absence of a negative sign problem is valid only
if the phonons are not integrated out, as done in Refs.~\cite{Assaad07,Chen19,Karakuzu18}. 
For example, the continuous-time interaction expansion (CT-INT) QMC method
\cite{Assaad07} suffers, in general, from a negative sign problem when
applied to two-dimensional (2D) electron-phonon problems. The approach introduced in
Ref.~\cite{Karakuzu18} for the Hubbard-Holstein model is free of a sign
problem but only part of the parameter space is accessible.
Even in the absence of a sign problem, a central 
challenge is to find an adequate sampling scheme that deals with  the
separation of energy scales.  Adopting a local updating scheme---as commonly
used in QMC simulations of fermions---in which the phonon field is
updated on a time scale set by the electron motion, leads to prohibitively
long autocorrelation times  \cite{Hohenadler08}. Global updates of the
phonon fields on the imaginary time scale of the inverse Debye temperature are
highly desirable and have been achieved using, for example, self-learning
methods \cite{Chen18}. Finally,
a sufficiently favorable scaling of the numerical effort with system size is
essential to study phase transitions.

\begin{figure}[b]
\includegraphics[width=0.6\linewidth]{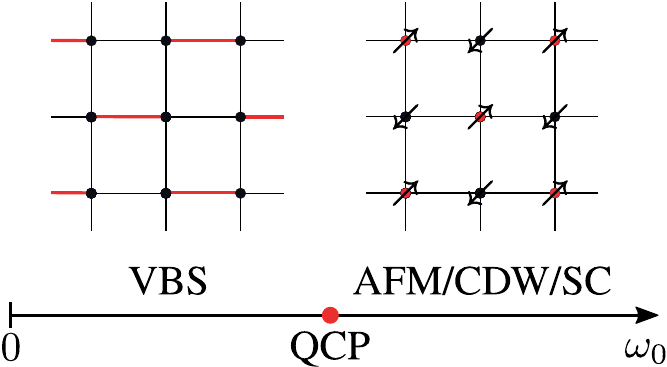} \\
 \caption{\label{Phase_diagram.fig} Schematic phase diagram of the
   O(4)-symmetric, square-lattice SSH model as a function of phonon
   frequency. The left inset depicts the VBS phase, showing one of four equivalent $(\pi,\pi)$ patterns of strong
     and weak bonds. The right inset illustrates the AFM phase which, due to the O(4)
     symmetry, is equivalent to a phase with CDW or SC order.}
\end{figure}

The history of unbiased numerical results for 2D
  systems of electrons coupled to quantum phonons directly reflects these
  algorithmic challenges. QMC investigations focused on the two fundamental
  types of
interactions captured by the Holstein \cite{HOLSTEIN1959325} and Su-Schrieffer-Heeger (SSH) Hamiltonians
\cite{Su80}, respectively. Simulations of Holstein-type models have a long
history, mainly in connection with CDW formation and superconductivity (SC), but many
important questions have only been resolved recently, see, e.g.,
Ref.~\cite{PhysRevB.100.165114} and references therein. QMC simulations
of the simplest variant of the SSH model, namely a half-filled square lattice,
have remained elusive until recently when the long-standing discussion regarding
the pattern of the expected VBS ground state was resolved in Ref.~\cite{Xing21}.

Here, we investigate the 2D SSH model with optical phonons on the square
  lattice and for the case of a half-filled band. We focus on the high-symmetry
  point with partial particle-hole symmetry that exhibits an O($2N$) symmetry
  for the general case of fermions with $N$ flavors. 
  
The purpose of our work is twofold.
First, we investigate the usefulness of Langevin updates,
successfully applied before to the 2D Holstein model \cite{Batrouni19}, by using
an identical algorithm based on the standard auxiliary-field QMC (AFQMC) formulation \cite{Blankenbecler81}
implemented in the ALF-2.0 package \cite{ALF_v2}. Our analysis is
based on the calculation of a Pfaffian whose sign changes track the zeros of the
fermion determinant. Following Ref.~\cite{Batrouni19}, we also
implemented Fourier acceleration to reduce autocorrelations. Our analysis
reveals that the method is suited to investigate the particularly interesting
adiabatic regime.

Second, we provide insight into the physics of the SSH model, namely
evidence for and details of a transition from a VBS
state to an AFM state with increasing phonon frequency.
A schematic phase diagram is shown in Fig.~\ref{Phase_diagram.fig}. 
AFM order at finite phonon frequencies is remarkable as it was expected
only in the presence of Coulomb repulsion \cite{PhysRevB.37.9546,PhysRevB.46.1710,YuanKopp2001,PhysRevB.65.085102,JPSJ.73.2777}.
We characterize the evolution with phonon frequency by calculating
susceptibilities and, in particular, excitation spectra. We complement our
numerical results with those from a self-consistent Born
approximation. Furthermore, we determine the mean-field ground state
and study its destruction by thermal fluctuations in the
adiabatic limit.

The rest of the article is organized as follows. In
Sec.~\ref{Symmetries.sec}, we define the SSH model and comment on symmetries,
limiting cases, and previous work. In Sec.~\ref{Methods.sec}, we
discuss the numerical method. Our results are presented in
Sec.~\ref{Results.sec}, followed by a discussion in Sec.~\ref{Summary.sec}. We
also provide an appendix with details about the self-consistent Born approximation.

Aspects of this work were already reported in Ref.~\cite{Beyl_thesis}.

\section{Model and Symmetries}\label{Symmetries.sec}

\subsection{Hamiltonian}

We consider an SSH model with optical phonons, defined by the Hamiltonian
\begin{eqnarray}\label{eq:SSH}
\hat{H}_{\text{el}} =& -& t   \sum_{\langle \boldsymbol{i}, \j \rangle}   \sum_{\sigma =1}^{N}
	  \left( \hat{c}^{\dagger}_{\i,\sigma}
            \hat{c}^{\phantom{\dagger}}_{\j,\sigma}    + \text{h.c.} \right) + \sum_{b} \left[ \frac{\hat{P}^2_b}{2 m} + \frac{k}{2} \hat{Q}_b^2 \right]  \nonumber\\ 
&+& g  \sum_{\langle \i, \j \rangle}    \hat{Q}_b  \sum_{\sigma =1}^{N}
	  \left( \hat{c}^{\dagger}_{\i,\sigma}
            \hat{c}^{\phantom{\dagger}}_{\j,\sigma}    + \text{h.c.} \right). 
\end{eqnarray}
The first term describes the hopping of electrons on the  bonds $b=\langle \i,\j \rangle$ 
connecting two nearest-neighbor sites $\i$, $\j$ with hopping amplitude $t$.
The operator $\hat{c}_{\i,\sigma}^{\dagger}$ creates an electron in a  Wannier  state centered at
site $\i$ and with $z$-component of spin $\sigma$ that runs over $N$
flavors. We use anti-periodic
boundary conditions  $\hat{c}_{\i+L \bm{a}_1,\sigma}^{\dagger} = -\hat{c}_{\i,\sigma}^{\dagger}$
 in the direction of the primitive vector $\bm{a}_1$ of the lattice and 
periodic boundary conditions  $\hat{c}_{\i+L
  \bm{a}_2,\sigma}^{\dagger} = \hat{c}_{\i,\sigma}^{\dagger}$ in the
direction of $\bm{a}_2$. The phonons are represented by harmonic oscillators that
reside on the bonds. They are described by momentum and position operators
$\hat{P}_b$ and $\hat{Q}_b$ as well as the frequency $\omega_0^2={k}/{m}$, 
where $k$  is the spring constant and $m$ the mass of the harmonic oscillators.
Electron hopping is modulated by the coupling to the phonon coordinate
$\hat{Q}_b$ on the respective bond $b$ with strength $g$. The adiabatic
regime is defined as $\omega_0<t$. All results will be for $N=2$ (\ie, spin-1/2 electrons).

\subsection{Symmetries}

The SSH model at half-filling and on a bipartite lattice is invariant under the
partial particle-hole transformation
\begin{eqnarray}\label{eq:PHT}
\hat{P}^{-1}_{\sigma} \hat{c}_{\i,\sigma'}^{\dagger} \hat{P}_\sigma =
  \delta_{\sigma,\sigma'} \text{e}^{\text{i} \bm{Q} \cdot \i }
  \hat{c}_{\i,\sigma'}^{\phantom{\dagger}} + \left(1-\delta_{\sigma,\sigma'}
  \right) \hat{c}_{\i,\sigma}^{\dagger}
\end{eqnarray}
where $\bm{Q}  = (\pi,\pi)$ for the square lattice considered here.  
We can define a corresponding $\mathbb{Z}_2$ order parameter, the fermion
parity on site $\i$ \cite{Assaad16},
\begin{eqnarray}\label{eq:par}
\hat{p}_{\i} = \prod_{\sigma=1}^{N} \left( 1- 2\hat{n}_{\i,\sigma} \right) .
\end{eqnarray}
This Ising-like order parameter supports order at finite temperature. Since it
changes sign under transformation~(\ref{eq:PHT}), it can be used to detect a
spontaneous breaking of the particle-hole symmetry.

In addition to the apparent global $\mathrm{SU}(N)$ spin rotation symmetry,
the model possesses an enlarged $\mathrm{O}(2N)$ symmetry on a bipartite
lattice. To prove this, we reformulate the Hamiltonian using Majorana
fermions \cite{Assaad16,Beyl17}
\begin{eqnarray}
\hat{c}_{\i,\sigma}^{\dagger} = \frac{1}{2} \left( \hat{\gamma}_{\i,\sigma,1} - \text{i} \hat{\gamma}_{\i,\sigma,2} \right).
\end{eqnarray}
After a canonical transformation $\hat{c}_{\i}^{\dagger} \rightarrow \text{i}
\hat{c}_{\i}^{\dagger}$ on one sublattice, the hopping operator can be written as
\begin{eqnarray}\label{eq:Kb}
\hat{K}_b =   \sum_{\sigma =1}^{N}
	  \left( \hat{c}^{\dagger}_{\i,\sigma}
            \hat{c}^{\phantom{\dagger}}_{\j,\sigma}    + \text{h.c.} \right) = \frac{\text{i}}{2} \sum_\sigma \sum_{\alpha=1}^2  \hat{\gamma}_{\i,\sigma,\alpha} \hat{\gamma}_{\j,\sigma,\alpha},
\end{eqnarray}
thereby revealing the $\mathrm{O}(2N)$ symmetry. Because of the latter,
the model is free of a sign problem for odd values of $N$
\cite{Li16}. For even $N$, time-reversal symmetry is sufficient to show the
absence of a sign problem \cite{Wu04}. 
In the case of $N=2$ considered here, the spin operators and the
Anderson pseudospin operators \cite{Anderson58} are the infinitesimal
generators of the $\mathrm{SO}(4)$ symmetry. They are defined by
\begin{eqnarray}
\hat{\bm{S}}_{\i}&= \frac{1}{2} \sum_{\sigma,\sigma'} \hat{c}_{\i,\sigma}^{\dagger} \bm{\sigma}_{\sigma,\sigma'} \hat{c}_{\i,\sigma'}, \quad 
\hat{\bm{\eta}}_{\i} &= \hat{P}_{\uparrow}^{-1}\hat{\bm{S}}_{\i}\hat{P}_{\uparrow},
\end{eqnarray}
where the vector $\bm{\sigma}$ contains the three Pauli matrices.
The spin and pseudospin components ($l$, $m$, $n$) fulfill the Lie
algebra of the $\mathrm{SU}(2)$ group $\big[\hat{S}_{\i,l}, \hat{S}_{\j,m}
\big] = \text{i}  \delta_{\i,\j} \sum_n \varepsilon_{lmn} \hat{S}_{\i,n}$
and commute among each other. Here, $\varepsilon_{lmn}$ is the Levi-Civita
symbol. The Lie algebra of the global $\mathrm{O}(4)$ symmetry can be
interpreted as $\mathrm{O}(4)=\mathrm{SU}(2)\times \mathrm{SU}(2) \times \mathbb{Z}_2
$,  where the additional $\mathbb{Z}_2$ symmetry corresponds to the partial
particle-hole symmetry \cite{Assaad16}.   Hence, an AFM
phase is degenerate with a CDW and an s-wave
superconductor (SC). If the parity $\hat{p}_{\i}$ orders and the
particle-hole symmetry is  spontaneously  broken, either the spin or charge
sector is explicitly chosen.
The VBS ground state in Fig.~\ref{Phase_diagram.fig} spontaneously breaks the
$C_4 $ symmetry of the lattice, whereas long-range AFM order breaks the O(4) symmetry down to U(1).

\subsection{Limiting cases}

In the adiabatic limit $\omega_0 \rightarrow 0$, imaginary-time fluctuations
of the phonon fields are exponentially suppressed.  The phonon displacements can be
treated classically and the Hamiltonian can be written as
\begin{align}\label{eq:adiabaticH}
\hat{H}= \sum_b (-t + g q_b) \hat{K}_b +\sum_b q_b^2.
\end{align}
Here, $\hat{Q}_b | q \rangle = q_b |q \rangle$ and $\hat{K}_b$ is defined in Eq.~(\ref{eq:Kb}). The Hamiltonian consists
solely of a modulated hopping of the electrons and the potential energy of
the phonon fields. Mean-field theory or Monte Carlo simulations yield
a VBS ground state, see Sec.~\ref{sec:mean-field}.

For $\omega_0>0$, we can  integrate out the phonons to obtain an effective Hamiltonian for the
electrons \cite{Negele,Assaad16}. This yields the action
\begin{eqnarray}
S_{\text{eff}} 
&= -\frac{g^2}{2k} \int_0^{\beta} \int_0^{\beta} \text{d}\tau \, \text{d}\tau' \sum_{b} \hat{K}_{b}(\tau) D(\tau-\tau')  \hat{K}_{b}(\tau') 
\end{eqnarray}
with $\beta=T^{-1}$ the inverse temperature (we set $k_{\text{B}}=1$). The
interaction is local but retarded, 
\begin{align}
D(\tau)= 
 \frac{\omega_0}{2} \frac{\text{e}^{-\omega_0 |\tau|}+\text{e}^{-\omega_0 (\beta-|\tau|)}}{1-\text{e}^{-\omega_0 \beta }}.
\end{align}
By taking the antiadiabatic limit $\omega_0 \rightarrow \infty$, it becomes instantaneous \cite{Assaad16},
\begin{align}
\lim\limits_{\omega_0 \to \infty} D(\tau) = \delta(\tau).
\end{align}
The effective Hamiltonian of the SSH model in the antiadiabatic limit is
given by
\begin{align}
\hat{H}_{\text{eff}} = -  t  \sum_{\langle \i, \j \rangle}   \hat{K}_b
	   -\frac{g^2}{2k} \sum_{\langle \i, \j \rangle}  \hat{K}_b^2.
\end{align}
For $N=1$, this expression is equivalent to the Hamiltonian of the $t$-$V$
model if we set $g=\sqrt{kV}$. For two fermion flavors, $N=2$, we can rewrite the interaction term as
\begin{align}
\label{Kb2.eq}
- \frac{1}{4}  \hat{K}_b^2 =  \hat{\boldsymbol{S}}_{\i} \cdot  \hat{\boldsymbol{S}}_{\j} + \hat{\boldsymbol{\eta }}_{\i}  \cdot \hat{\boldsymbol{\eta}}_{\j}\,.
\end{align}
The interaction~(\ref{Kb2.eq}) again reveals the $\mathrm{O}(4)$ symmetry and
favors an AFM/CDW/SC ground state \cite{Assaad16}.

\subsection{Previous work}

Despite its long history \cite{Su80}, the SSH model was
mainly studied in connection with 1D materials. Until recently, investigations
of the 2D SSH model relied on mean-field arguments or started outright from the
adiabatic limit of classical phonons. The correct mean-field VBS pattern remained controversial
\cite{Mazumdar87,PhysRevB.37.9546,PhysRevB.39.12324,PhysRevB.39.12327} and an
alternative, multi-mode Peierls state with no well-defined ordering wavevector
was suggested \cite{JPSJ.69.1769,JPSJ.72.1995,JPSJ.73.2473}. AFM order in 2D SSH
models with additional Coulomb interaction was discussed in
Refs.~\cite{PhysRevB.37.9546,PhysRevB.46.1710,YuanKopp2001,PhysRevB.65.085102,JPSJ.73.2777}.
However, all these works completely left out the impact of quantum lattice
fluctuations, which are the focus of the present work and have proven to have a
crucial impact for the 1D spinless (\ie, $N=1$) SSH model \cite{Weber19}. Numerical
confirmation of the existence of a unique VBS ground state and its ordering
pattern was provided by QMC simulations in Ref.~\cite{Xing21}, where a nonzero
critical value was reported for quantum phonons. QMC results were also obtained
for the honeycomb lattice \cite{pub.1051655427} and the Lieb lattice \cite{li2020quantum}.

\section{Methods}\label{Methods.sec}

\subsection{Langevin dynamics}

Using the real-space formulation of the path integral for the phonon degrees
of freedom with the eigenstates $|q\rangle$ of the position operator
$\hat{Q}_b$, the partition function of the model
can be written as
\begin{eqnarray}\label{eq:S_SSH}
  Z & = &\int \prod_{b,\tau} \text{d}q_{b,\tau} \, \text{e}^{-S}, \\
  S &= & S_0 +S_{\text{F}} = S_0 - N \ln{ \det\left[\mathds{1}  + B(\beta,0) \right]} \nonumber
\end{eqnarray}
with
\begin{eqnarray}\label{eq:B_matrix}
  B(\tau_1,\tau_2) =\!\!\!\! \prod_{\tau=\tau_2+\Delta \tau}^{\tau_1}  \left(\prod_b \text{e}^{- \Delta \tau g q_{b,\tau} K_b} \right) \text{e}^{\Delta \tau t\sum_b  K_b}
\end{eqnarray}
and the matrix 
\begin{equation}
\left(K_b\right)_{x,y} = \begin{cases} 
1 &  \text{if $x\in b$ $\land$ $y\in b$ }\\
0 & \text{otherwise}
\end{cases}\,.
\end{equation}
Here, $x$ and $y$ label lattice sites; $x\in b$ means that site $x$ belongs to bond $b$.
In the path integral, we discretized the imaginary time interval $[0,\beta[$ into steps of width $\Delta\tau=\beta/L_{\text{Trot}}$.
Following Blankenbecler, Scalapino, and Sugar (BSS) \cite{Blankenbecler81} we
rewrote the fermionic trace as a determinant. Therefore, we only have to sample the phonon 
degrees of freedom, whose $g=0$ imaginary-time dynamics is governed by
\begin{eqnarray}
  S_0 =  \Delta \tau \sum_{b,\tau} \left( \frac{1}{\omega_0^2} \left[ \frac{q_{b,\tau+1}-q_{b,\tau}}{\Delta \tau} \right]^2 +  q_{b,\tau}^2 \right). 
  \label{eq:ssh_s0}
\end{eqnarray}
We use Langevin dynamics to update the phonon fields $\bm{q}=\{q_{b,\tau}\}$.
The corresponding Langevin equation is a stochastic differential
equation for the fields \cite{Batrouni19,ALF_v2},
\begin{eqnarray}
\label{eq:Langevin}
  \frac{\text{d}\bm{q}(t_l )}{\text{d} t_l}    =     - M \frac{\partial S(
  \bm{q}(t_l)) }{\partial    \bm{q}(t_l) }       +\sqrt{2  M} \bm{\eta}(t_l),
\end{eqnarray}
with an additional Langevin time $t_l$. The independent Gaussian random variables $\bm{\eta}$ satisfy 
\begin{align}
  \langle \eta_{b,\tau}(t_l) \rangle = 0, \quad
  \langle \eta_{b,\tau}(t_l) \eta_{b',\tau'}(t_l') \rangle  = \delta_{b,b'} \delta_{\tau,\tau'} \delta(t_l-t_l')\,,
  \label{eq:random}
\end{align}
where $\delta$ is to be understood as a Kronecker $\delta$ for the discrete indices and
as the Dirac $\delta$ function for the continuous $t_l$. The matrix $M$ is an arbitrary
positive-definite matrix.  In order to use the Langevin equation in our AFQMC
code, we discretize the Langevin time $t_l$ with a finite time step $\delta
t_l$. Using the Euler method, the discretized equation is given by \cite{Gardiner09}
\begin{align}\label{eq:langevin}
  \bm{q}(t_l +  \delta t_l)    =    \bm{q}(t_l)    - M \frac{\partial S( \bm{q}(t_l)) }{\partial    \bm{q}(t_l) }    \delta t_l     +\sqrt{2 \delta t_l M} \bm{\eta}(t_l).
\end{align}
For the random variables $\bm{\eta}$ we replace
$\delta(t_l-t_l') \rightarrow \delta_{t_l,t_l'}$. The systematic error
introduced by discretizing $t_l$ is of linear order in
$\delta t_l$ \cite{Batrouni19,ALF_v2}.  By transforming the Langevin equation
into a Fokker-Planck equation one can show that the stationary
probability distribution of finding the system in state $\bm{q}$ is given by
\cite{Batrouni85}
\begin{eqnarray}\label{eqn:langevin_solution}
  P(\bm{q}) =  \frac{ \text{e}^{ - S(\bm{q}) } }   {   \int D \bm{q} \, \text{e}^{ - S(\bm{q}) } }.
\end{eqnarray}

A major aspect of Langevin dynamics are the forces, their computation, and
characteristics.  For the SSH model, using Eq.~(\ref{eq:S_SSH}), the forces read
\begin{eqnarray}
  \label{eq:forces_SSH}   
  \frac{\partial S}{\partial q_{b,\tau}} &= & \Delta \tau k q_{b,\tau} + \frac{m}{\Delta \tau} \left( 2 q_{b,\tau} -q_{b,\tau+1} -q_{b,\tau-1} \right)  \\
                                         & \phantom{=} &+ N g \Delta \tau \Tr{ \left\{K_b (1-G(b,\tau)) \right\}}  \nonumber                                                
\end{eqnarray}
with the Green function
\begin{eqnarray}\label{eq:S_Green}
  G_{\i,\j}(b,\tau) = \frac{ \Tr\left[\hat{U}^{<}(b,\tau)  \hat{c}_{\i}^{\phantom{\dagger}} \hat{c}_{\j}^{\dagger} \hat{U}^{>} (b,\tau)\right]}{\Tr\left[\hat{U}(\beta,0) \right] }
\end{eqnarray}
and the propagators
\begin{eqnarray}\label{eq:U_Green}
  \hat{U}^{<}(b',\tau) &=&  \hat{U}(\beta,\tau) \prod_{b=b'}^{N_b} \text{e}^{-\Delta \tau g q_{b,\tau}\hat{\boldsymbol{c}}^{\dagger} K_b \hat{\boldsymbol{c}}},  \\
  \hat{U}^{>}(b',\tau) &= &\prod_{b=1}^{b'-1} \text{e}^{-\Delta \tau g q_{b,\tau}\hat{\boldsymbol{c}}^{\dagger} K_b \hat{\boldsymbol{c}} }
                            \text{e}^{\Delta \tau t \sum_b  \hat{\boldsymbol{c}}^{\dagger} K_b \hat{\boldsymbol{c}}} 
                            \hat{U}(\tau-\Delta\tau,0). \nonumber
\end{eqnarray}
Here, $N_b$ is the total number of bonds on the lattice. From Eq.~(\ref{eq:S_SSH}) we see that the action has logarithmic
divergences if the determinant vanishes. The $\mathrm{O}(2N)$ symmetry of the
model only guarantees that the determinant is non-negative.
An advantage of Langevin dynamics is that it amounts to global updates. In
each step, all phonon fields are updated and, contrary to the
Metropolis-Hastings algorithm \cite{Hastings1970,Krauth2006}, there is no
acceptance-rejection step.

To control the motion through configuration space we used an
adaptive Langevin time step $\delta t_l$ \cite{White88a,Assaad90}. At each Langevin time step, the
fermionic forces for every $b$ and $\tau$ are compared to a preset maximal
force $F_{\text{max}}$. If the maximal computed force
$\text{max}(\partial S_F/ \partial q_{b,\tau})$ exceeds
$F_{\text{max}}$, $\delta t_l$  is decreased by the ratio of the two forces,
\begin{align}
\bar{\delta t_l}= \frac{F_{\text{max}}}{\text{max}\left(\frac{\partial S_F} {\partial q_{b,\tau}} \right)} \delta t_l.
\label{eq:t_l}
\end{align}
The variations of the Langevin time step have to be accounted for when measuring observables,
\begin{align}
\langle \hat{O} \rangle =
  \frac{\sum_{\alpha=1}^{N_m} \left(\bar{\delta t_l} \right)_{\alpha}
  \langle\langle \hat{O}\rangle\rangle_{\alpha}}{\sum_{\alpha=1}^{N_m} \left(\bar{\delta t_l}
  \right)_{\alpha}}.
  \label{eq:measure}
\end{align}
Here, $N_m$ is the total number of measurements and $\langle
\langle \hat{O} \rangle \rangle_{\alpha}$ denotes the value of the observable
$\hat{O}$ for configuration $C_\alpha$ of the phonon
fields. Equation~(\ref{eq:measure}) reduces to
$\langle \hat{O} \rangle = \frac{1}{N_m} \sum_{\alpha=1}^{N_m} \langle\langle \hat{O}\rangle\rangle_{\alpha}$
for a fixed time step and to
$
  \langle \hat{O} \rangle = \frac{1}{T_l} \int_0^{T_l} \mathrm{d}t_l \langle
  \langle \hat{O}(t_l) \rangle \rangle
  $
for continuous Langevin time.

\begin{figure}
 \includegraphics[width=\linewidth]{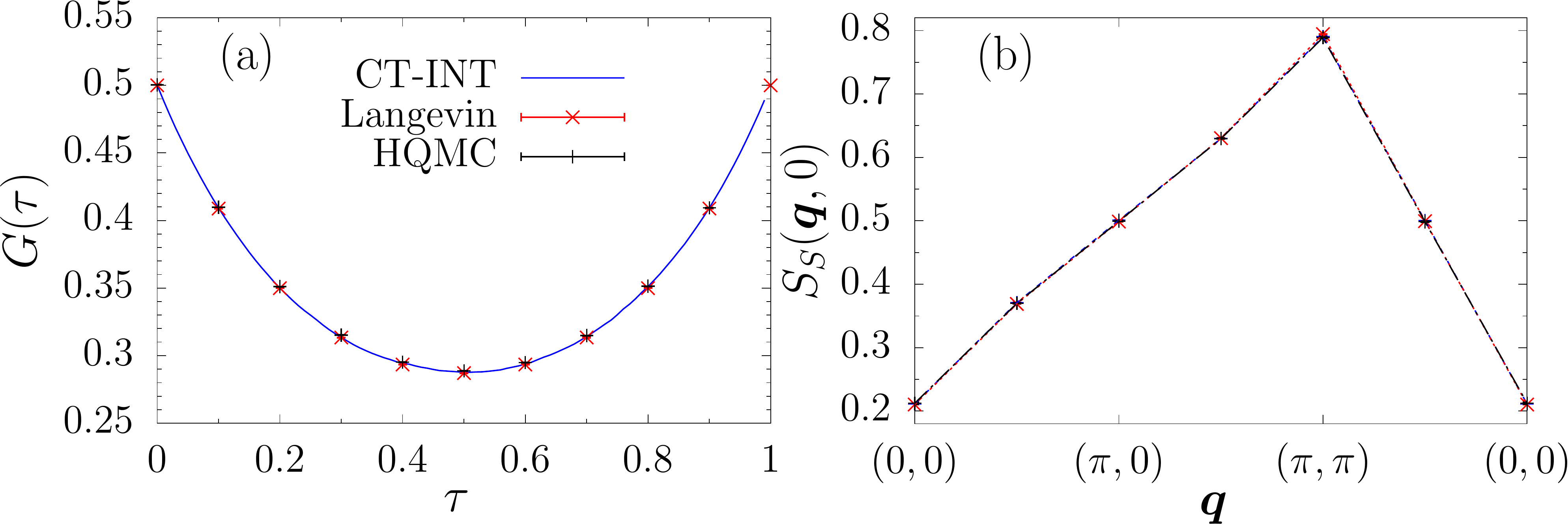}
  \caption{\label{fig:Bench} Comparison of Langevin, CT-INT and HQMC results for
    (a) the local, imaginary-time Green function and (b) the spin correlation
    function on an $L=4$ lattice for $\beta=1.0$, $\Delta \tau=0.1$, $g=1.0$,
    $k=2$, $\omega_0=1.0$, and $\delta t_l = 0.0005$. HQMC and CT-INT data taken from 
    Ref.~\cite{Beyl17}. }
\end{figure}

Simulations start from a random field configuration $\{q_{b,\tau}\}$ and
iterate the following set of steps:
\begin{enumerate}
\item   Compute the equal-time  Green functions on all time slices from Eq.~(\ref{eq:S_Green}). 
\item   Compute the forces via  Eq.~(\ref{eq:forces_SSH}).
\item   Using the equal-time Green functions from (1), we can compute $\langle\langle \hat{O}\rangle\rangle_{\alpha}$ 
for any equal-time,  multi-point correlation function, see
Eq.~(\ref{eq:measure}). To this end, we make use of Wick's theorem, which holds for a given field configuration.
\item   Adjust the time  step according to  Eq.~(\ref{eq:t_l}).
\item   Draw a set of independent Gaussian variables $\bm{\eta}$.
\item   Compute new  fields $\{q_{b,\tau}\}$ from the Langevin equation using the Fourier acceleration
  matrix $M$ and running  Langevin time step  $\bar{\delta t_l}$ [Eq.~(\ref{eq:Langevin})].
\end{enumerate}  

In Fig.~\ref{fig:Bench}, we compare selected results from our method with two
other QMC approaches: the hybrid QMC (HQMC) method described for the SSH
model in Ref.~\cite{Beyl17} and CT-INT QMC method in which the phonons are
integrated out in favor of a retarded interaction of the electrons
\cite{Rubtsov05,Assaad07,Weber15a}. The results from all three methods are in
good agreement.

\subsection{Calculation of the Pfaffian}

In this section, we explore how suitable Langevin
  dynamics is for the SSH model and in which parameter regions it is
  particularly efficient or problematic.  As  opposed  to the HQMC approach of
  Ref.~\cite{Beyl17}, Langevin updates constitute   rejection-free  global  moves. 
  A key requirement for
the success of these methods is the \textit{absence} of
singularities in the action $S$.  Since $S$ contains the logarithm of the
fermion determinant, the latter must not vanish. In special cases where $S$ has no
singularities, this class of updating schemes works very well. A notable example is
the 1D Hubbard model with open boundary conditions
\cite{ALF_v2}. For the SSH model in the adiabatic limit, the phonon fields are
frozen in imaginary time and the fermion determinant is strictly positive. For
$\omega_0>0$, this is not the case.

To analyze the fermion determinant, we derive a relation to a Pfaffian. The latter can be evaluated
numerically and its sign changes track the zeros of the determinant. Note that
the Pfaffian is not necessary for simulations but rather represents an additional diagnostic.

The $\mathrm{O}(2N)$ symmetry of the model permits
us to express the determinant as a square of a trace over one of the two Majorana fermions:
\begin{eqnarray}
  Z_{\gamma}^2 &=& \det\left[\mathds{1}  + B(\beta,0) \right]  \\
  Z_{\gamma} &=&  \Tr \left[  \prod_\tau \prod_b \left( \text{e}^{- \frac{\text{i}}{2} \Delta \tau g q_{b,\tau}  \hat{\gamma}_{\i} \hat{\gamma}_{\j}}  \right)
                 \text{e}^{\frac{\text{i}} {2} \Delta\tau t\sum_b  \hat{\gamma}_{\i} \hat{\gamma}_{\j}} \nonumber
                 \right]. 
\end{eqnarray}
Here and in the rest of this section, we drop the spin and Majorana kind
indices since none of the quantities considered explicitly depend on them.
One can show with a canonical transformation of the Majorana fermions on only
one sublattice, $ \hat{\gamma}_{\i} \rightarrow - \hat{\gamma}_{\i} $, that
$Z_\gamma$ is real \cite{Yao14a} and its square hence non-negative.

$Z_{\gamma}$ can have a different sign in different regions of the configuration
space. Being an entire function, it necessarily has to vanish between these regions. Hence, the average sign
of $Z_\gamma$ serves as an estimate of the number of zeros of the determinant.
If the average sign is close to plus or minus unity, we are less likely
to cross a boundary between two regions in which $Z_\gamma$ has
different signs. In contrast, a small average sign implies more zeros.

To measure the sign of $Z_\gamma$ we reformulate it as a
Pfaffian. First, we use an alternative Trotter decomposition and rewrite
the exponentials as hyperbolic functions by using
$\left( \hat{\gamma}_{\i} \hat{\gamma}_{\j}\right )^2 =-1$ to obtain
\begin{eqnarray}
  Z_{\gamma} &=& \Tr{ \left[ \prod_{x}  \text{e}^{  \text{i} y_{x}   \hat{\gamma}_{\i} \hat{\gamma}_{\j}  }  \right] } \\
             &=&  \prod_{x} \left(\cosh y_{x}\right) \Tr{ \left[ \prod_{x} \left(1 + \text{i} \hat{\gamma}_{\i} \hat{\gamma}_{\j} \tanh{y_{x}}   \right) \right] } .
                 \label{eq:Majorana_Tr}\nonumber
\end{eqnarray}
The tuple $x=(b,\tau)$ combines the bond index and the imaginary time slice
into a new index ordered according to its position in
the product $\prod_\tau \prod_b$. To lighten the notation, we used 
$y_{x}=\frac{1}{2}\Delta \tau \left( t - g q_{x} \right)$. Next,
we introduce Grassmann variables $\xi_{\i/\j,\tau}$ \cite{Negele} on every site and
imaginary time slice, where $\i$ and $\j$ are on different sublattices,
and use
\begin{eqnarray}\nonumber
C_{\pm}
 \prod_{x=1}^n\sqrt{ a(x)} &=& \int \left[ \text{d} \xi \right] \text{e}^{\pm \sum_{x<x'} \sqrt{a(x) a(x')} \xi_{x'} \xi_x}, \\
 C_+ &=& (-\text{i})^n,  \quad C_-=1
\end{eqnarray}
for even $n$ \cite{Huffman18a}.
Here, $ \left[ \text{d} \xi \right] =  \text{d} \xi_{n}...\text{d}\xi_1$  is a
time-ordered product and $a\in\mathbb{C}$. Finally, $Z_\gamma$ can be
written as the Pfaffian over an antisymmetric matrix $A \in
\mathbb{C}^{2N_bL_{\text{Trot}}\times 2 N_b L_{\text{Trot}}}$ \cite{Huffman18a},
\begin{equation}\label{eq:Pfaff1}\nonumber
 Z_{\gamma}   =  \prod_{x} \left(\cosh{y_{x}}\right) \Tr{(1)} \Pf(A)\,,
\end{equation}
where
\begin{equation}\label{eq:Pfaff2}
\Pf(A) = \int  \left [ \text{d} \xi \right] \text{e}^{-\frac{1}{2}\bm{\xi}^T A \bm{\xi}}
\end{equation}
and
\begin{eqnarray}\label{eq:Pfaff3}\nonumber
-\frac{1}{2}\bm{\xi}^T A \bm{\xi} &=& -\sum_x \xi_{\i,x} \xi_{\j,x} +
  \sum_{\i,(x<x')} m_{x'x} \xi_{\i,x'} \xi_{\i,x} \\
  &\phantom{=}& - \sum_{\j,(x<x')} m_{x'x} \xi_{\j,x'} \xi_{\j,x}\,. 
\end{eqnarray}
Here, $m_{x'x} = \sqrt{ \tanh{(y_{x})} \tanh{(y_{x'})}}$ and the vector $\bm{\xi}$ contains all Grassmann variables. 

\begin{figure}
\includegraphics[width=\linewidth]{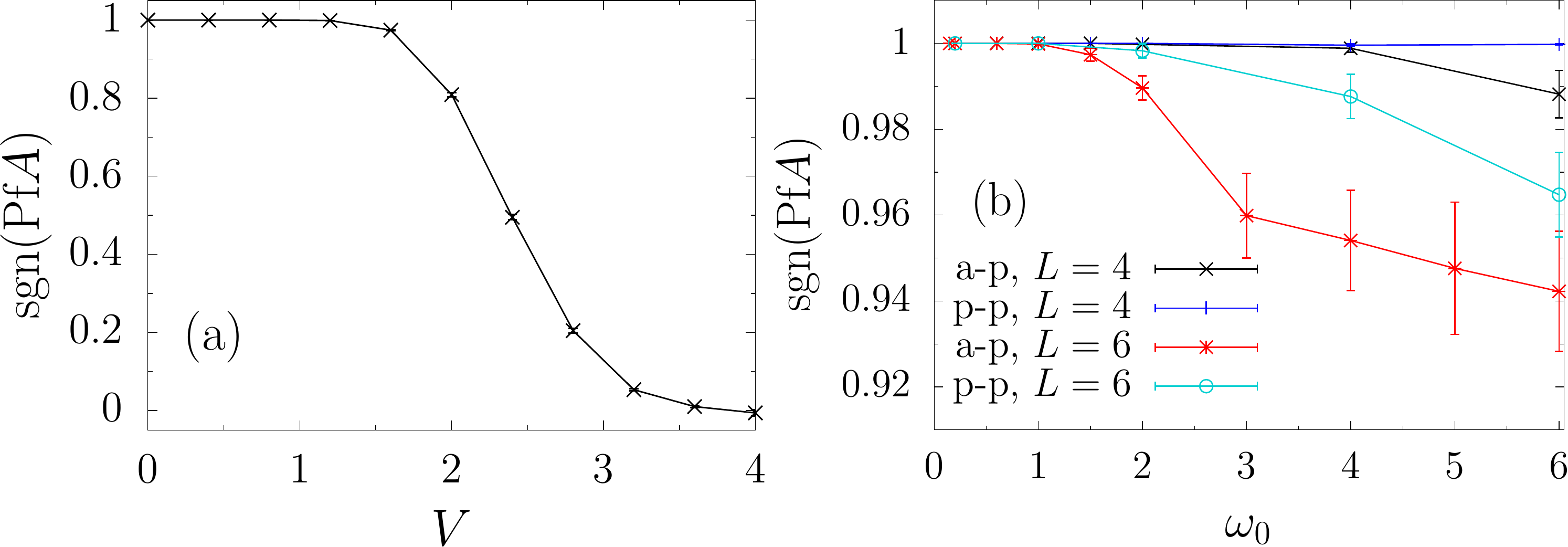}
  \caption{\label{fig:Pfaff} (a) Average sign of the Pfaffian for the
    $t$-$V$ model on the $\pi$-flux square lattice with $L=4$, $\beta t=4$, and
    $\delta t_l=0.005$. This model exhibits a Gross-Neveu phase 
    transition at $V_c = 1.279(3)t$  \cite{Huffman19}.  (b) Average sign of the
    Pfaffian for the SSH model with antiperiodic-periodic (a-p) or
    periodic-periodic (p-p) boundary conditions and $\beta=5.0$,
    $\Delta \tau  = 0.1$, and $\delta t_l = 0.0005$.
  }
\end{figure}

For the numerical computation of the Pfaffian we used the software from
Ref.~\cite{Wimmer12}. Since the calculation is very
expensive, we only considered small lattices and a small number of imaginary
time slices.  In Fig.~\ref{fig:Pfaff},  we plot the average sign
for both the t-V model as a function of the interaction strength $V$ and 
the SSH model as a function of the phonon frequency $\omega_0$.

The results for the t-V model illustrate a breakdown of Langevin dynamics
due to severe divergences of the forces. Upon increasing $V$, the average sign of the Pfaffian drops to zero and
the measured observables deviated by up to a factor of $10^6$ from results
obtained with a Metropolis-Hastings updating scheme. In the adiabatic limit of the
SSH model, the average sign is close to unity and the simulations are
stable. Increasing $\omega_0$ leads to a decrease of the average sign
and the updating tends to become unstable. In general, we found it easier to
stabilize the simulations with our choice of mixed boundary conditions as
opposed to periodic boundary conditions in both directions.

\subsection{Fourier acceleration}

Following Refs.~\cite{Batrouni85,Batrouni19}, we used Fourier acceleration
to reduce autocorrelations.  Its main idea is to increase (reduce) the step size of the Langevin time of
slow (fast) phonon modes by using an adequate choice of the matrix $M$ in the Langevin
equation~(\ref{eq:langevin}) \cite{Davies86}. 

As a foundation for the choice of $M$ we consider the non-interacting case
($g=0$).  We carry out a Fourier transformation of the force in imaginary
time,
\begin{eqnarray}\label{eq:force_fourier}
  \hat{\bm{F}}\left[ \frac{\text{d}S}{\text{d} q_{b,\tau}} \right]
  =&\left[ \Delta \tau k  + \frac{2m}{\Delta \tau} \left( 1- \cos{\left( 2 \pi \nu_n \right) } \right) \right] q_{b,\nu_n}\,. 
\end{eqnarray}
To this end, we  defined the Fourier transformation for a function $f$ as
\begin{eqnarray}
  \hat{\bm{F}} \left[f(\tau) \right]  &  = \frac{1}{L_{\text{Trot}}} \sum_{\tau=1}^{L_{\text{Trot}}} \text{e}^{\text{i}2 \pi \nu_n \tau} f(\tau)
\end{eqnarray}
with
$\nu_n=\frac{n}{L_{\text{Trot}}}$ with $n=-\frac{L_{\text{Trot}}}{2}+1,
-\frac{L_{\text{Trot}}}{2}+2,...,\frac{L_{\text{Trot}}}{2}$. The ratio of the
slowest and fastest modes is \cite{Batrouni19}
\begin{eqnarray}
  \frac{(\Delta\tau)^2 k }{(\Delta\tau)^2 k + 4m } \ll 1.
\end{eqnarray}
Especially for $\omega_0\ll1$, it is close to zero. We choose the factor
$\tilde{M}(\nu_n)$ in Fourier space such that
the prefactor of $q_{b,\nu_n}$ in Eq.~(\ref{eq:force_fourier})
becomes independent of $\nu_n$,
\begin{eqnarray}
  \tilde{M}(\nu_n) = \frac{ \Delta \tau k+\frac{4m}{\Delta \tau}}{ \Delta \tau k+\frac{m}{\Delta \tau}\left( 2-2 \cos{\left( 2 \pi \nu_n  \right) } \right) }.
\end{eqnarray}
Although this choice is guided by the non-interacting case $g=0$, we also use
it for $g>0$ \cite{Batrouni19}. The modified Langevin equation reads
\begin{eqnarray}\label{eq:FA}
 \bm{q}(t_l+ \delta t_l) 
   = \bm{q}(t_l) &-& \hat{\bm{F}}^{-1} \left[ \delta t_l   \tilde{M}(\nu_n)
     \hat{\bm{F}} \left[ - \frac{\text{d}S}{\text{d}\bm{q} (t_l)}
     \right]\right.
  \\
  \quad
     &-&\left.
     \sqrt{2 \delta t_l}
     \sqrt{ \tilde{M}(\nu_n)} \hat{\bm{F}} \left[\bm{\eta}(t_l) \right] \right]. \nonumber
\end{eqnarray}

\begin{figure}[t]
  \includegraphics[width=\linewidth]{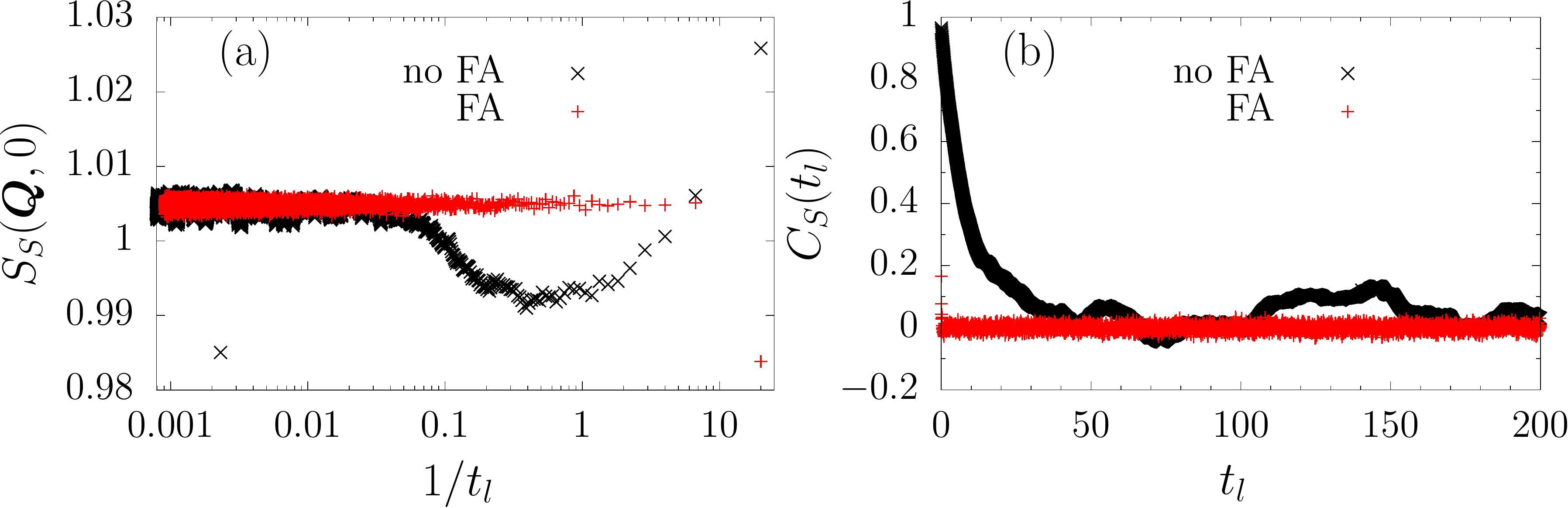}
  \caption{\label{fig:FA} (a) Equal-time spin correlation function
    $S_{S}(\bm{q},0)$ and (b) corresponding autocorrelation function
    $C_{S}(t_l)$ at wave vector $\bm{Q}$ as a function of Langevin
    time $t_l$ with and without the use of Fourier acceleration (FA). The 
    parameter sets for both runs are exactly the same (including number
    of sweeps/bins): $\omega_0 =0.4$, $L=6$, $\beta=7.0$, and 
    $\delta t_l=0.001$. 
    }
\end{figure}

To see the effect of Fourier acceleration on autocorrelations, we measured the
equal-time spin correlator
\begin{eqnarray}
  S_{S}(\bm{q},0)
  =\frac{1}{L^2} \sum_{\i,\j} && \text{e}^{-\text{i}\bm{q}(\i-\j)} \\
  %\hspace*{9em}
   &\phantom{=}&\times \left(\left \langle \hat{S}_{\i,z} \hat{S}_{\j,z} \right \rangle - \left\langle \hat{S}_{\i,z} \right \rangle \left\langle \hat{S}_{\j,z} \right \rangle \right)
      \nonumber
\end{eqnarray}
with [using Eq.~(\ref{eq:FA}) to update the fields] and without (by setting
$M=1$) Fourier acceleration.  Results at wave vector $\bm{Q}$ are shown in
Fig.~\ref{fig:FA}(a) as a function of the inverse Langevin time. 
The equilibration time is obviously reduced by Fourier acceleration and the
results of both methods agree at sufficiently long times.

We also consider the autocorrelation function \cite{ALF_v2}
\begin{equation}\label{eq:auto}
  C_{\hat{O}}(t_l) = \sum_{t'_l=0}^{T_l-t_l} \frac{\left(O(t'_l) -
  \langle \hat{O} \rangle \right)\left(O(t'_l+t_l) - \langle
  \hat{O} \rangle \right)}{\left(O(t'_l) - \langle \hat{O}
  \rangle \right)^2}.
\end{equation}
$O(t_l)$ is the observable evaluated at time $t_l$ and $T_l$ is the
maximal time at which measurements were taken. Shorter autocorrelation
times imply a faster decay of the autocorrelation function. A
decrease of autocorrelations by Fourier acceleration is clearly visible in Fig.~\ref{fig:FA}(b).

\section{Results}\label{Results.sec}

The key questions to be addressed are as follows. Starting from the exact
mean-field VBS ground state at $\omega_0=0$ (established in
Sec.~\ref{sec:mean-field}), what is the impact of thermal fluctuations?
What happens upon enhancing quantum lattice fluctuations by increasing
$\omega_0$ at fixed electron-phonon coupling? Does
the AFM order suggested by the interaction~(\ref{Kb2.eq}), derived for
$\omega_0=\infty$, emerge at finite and potentially experimentally relevant
phonon frequencies? Are the VBS and AFM phases connected by a single phase
transition or via an intermediate metallic phase? Finally, how does the
evolution from VBS to AFM order manifest itself in the spectral properties?

Simulations were done for spin-1/2 fermions ($N=2$) on $L\times L$ square lattices with mixed
boundary conditions, see Sec.~\ref{Symmetries.sec}.
We set $k=2$, $t=1$, $g=1.5$ and $\Delta \tau  = 0.1$. The Langevin time step
was $\delta t_l = 0.01$ for $\omega_0=0$ and $\delta t_l=0.0005$ else.

All simulations were carried out using the ALF package \cite{ALF_v2},
which provides a generic, high-performance implementation of the AFQMC method as well as
  tools for stochastic analytic continuation and error analysis.
Error bars were obtained using binning analysis
  \cite{Krauth2006,Gubernatis16,ALF_v2} and the delete-1 jackknife scheme
  \cite{Efron81,Gubernatis16,ALF_v2}, respectively.

\subsection{Adiabatic limit}\label{sec:mean-field}

\begin{figure}[t]
   \includegraphics[width=0.9\linewidth]{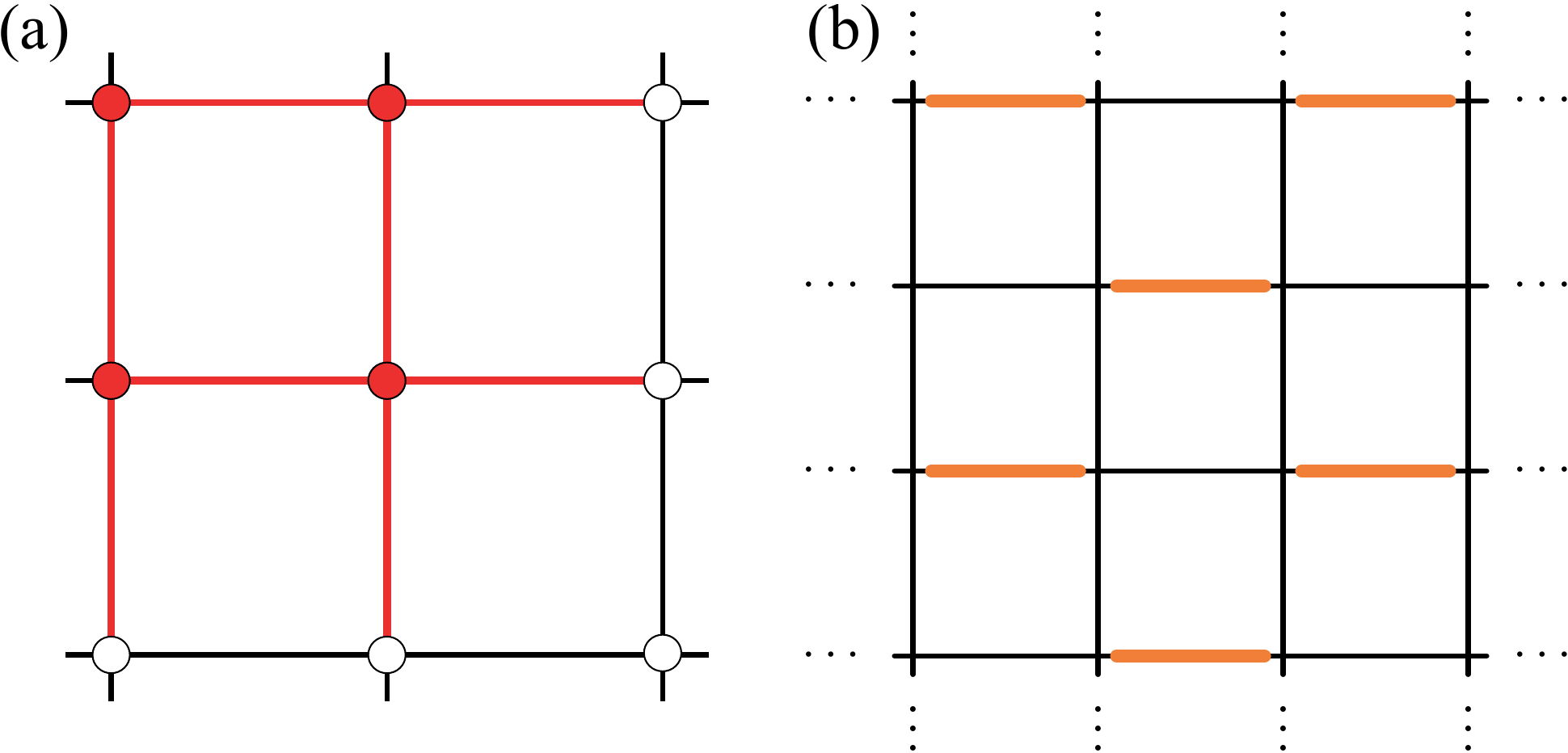}
  \caption{ \label{fig:pi-pi-VBS}(a) Unit cell used in mean-field theory,
    indicated by the colored sites and bonds. (b) Resulting minimal-energy 
    $(\pi,\pi)$ VBS pattern. Strong
    bonds are colored while weak bonds are represented by black lines.}
\end{figure}

For completeness, we use a mean-field approach to find the minimal energy
configuration of the classical fields $q_b$ and hence the ground state of Eq.~(\ref{eq:adiabaticH}).
The nested Fermi surface of the non-interacting problem gives rise to a log
divergence of the $\bm{q}=\bm{Q}$ bond susceptibility at low temperatures
and hence to a gapped VBS ground state. In contrast to one dimension, where the
ordering pattern is unique, possible 2D VBS patterns 
include staircase, 
columnar, staggered, and plaquette arrangements \cite{Mazumdar87,PhysRevB.37.9546}.
We use a $2\times2$ unit cell and vary the bond
variables independently according to the
aforementioned symmetry (see Fig.~\ref{fig:pi-pi-VBS}(a)), thereby allowing all
$(\pi,\pi)$ and $(0,\pi)$ patterns. Energy minimization yields the $(\pi,\pi)$
staggered VBS state illustrated in Fig.~\ref{fig:pi-pi-VBS}(b). The same pattern
was observed numerically in Ref.~\cite{Xing21}.

The phonons enhance the hopping amplitude on all bonds and effectively
renormalize the electronic bandwidth. Furthermore,
they modulate the hopping in a $(\pi,\pi)$ pattern (see
Fig.~\ref{fig:pi-pi-VBS}(b)) that leads to a finite gap at the Fermi
surface.  Since the VBS ordering breaks the discrete $\mathrm{C}_4$ symmetry
of the lattice, it can survive at finite temperatures.

To study the thermal melting of the VBS state, we consider
the partition function of Hamiltonian~(\ref{eq:adiabaticH}),
\begin{eqnarray}
  Z &=& \int \prod_b \text{d}q_b \, \text{e}^{-S}, \\
  S&=&  \beta \sum_b \frac{k}{2} q_b^2 -N_\sigma\ln{ \det{\left[ \mathds{1} + \text{e}^{-\beta\sum_b (-t + g q_b) K_b} \right]}} . \nonumber
\end{eqnarray}
The determinant is strictly positive since its argument is a symmetric
matrix. Therefore, we can use Langevin dynamics without divergences in the
forces to update the phonon fields via
\begin{eqnarray}
  \frac{\partial S}{\partial q_b} &=& \beta k q_b + \beta g N \Tr{\left\{ K_b \left( 1- G \right) \right\}}, \\
  G_{\i,\j} &=& \frac{\Tr{\left[ \text{e}^{-\beta \sum_b (-t + g q_b) \hat{K}_b
                } \hat{c}_{\i}^{\phantom{\dagger}} \hat{c}^\dagger_{\j}
                \right]}}{\Tr{\left[ \text{e}^{-\beta \sum_b (-t + g q_b)
                \hat{K}_b } \right]}}. \nonumber           
\end{eqnarray}
The onset of VBS order can be captured by the bond-kinetic susceptibility
\begin{equation}\label{eq:Kin_suscep}
  \chi_K^{\delta,\delta'} (\bm{q}) =   \int_{0}^{\beta}  \text{d} \tau \, S^{\delta,\delta'}_{K}  (\bm{q},\tau) 
\end{equation}
with the imaginary-time-displaced correlation function
\begin{eqnarray}\label{eq:Kin_corr}
  S^{\delta,\delta'}_{K}  (\bm{q},\tau) &=&
  \left< \hat{K}^{\delta}(\bm{q},\tau)\hat{K}^{\delta'}(-\bm{q}) \right>\\\nonumber
  &&\quad\quad- \left<  \hat{K}^{\delta}(\bm{q}) \right>   \left< \hat{K}^{\delta'}(- \bm{q}) \right>
\end{eqnarray}
and
\begin{equation}
  \hat{K}^{\delta}(\bm{q})   = \frac{1}{\sqrt{N}} \sum_{\bm{i},\sigma} e^{i
    \bm{q}\cdot \bm{i}}  \left(  \hat{c}^{\dagger}_{\bm{i},\sigma}
    \hat{c}^{\phantom\dagger}_{\bm{i} + \bm{a}_{\delta},\sigma}  +   \text{h.c.}
  \right).
\end{equation}

\begin{figure}[t]
  \includegraphics[width=0.49\linewidth]{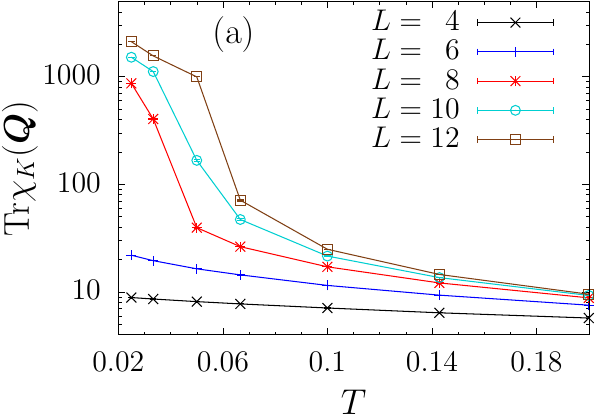}
  \includegraphics[width=0.49\linewidth]{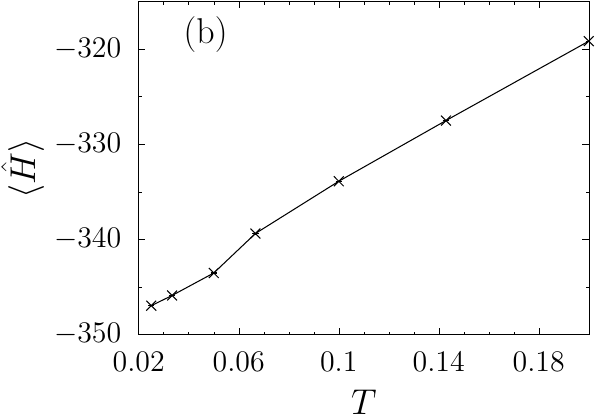}
 \caption{\label{fig:Mean-Field} (a) Bond susceptibility as a
   function of temperature for different system sizes. (b) Energy as a function of
   temperature for $L=12$.  Simulations were done using $\delta t_l =
   0.01$ and starting from the mean-field configuration.
 }
\end{figure}

Figure~\ref{fig:Mean-Field} shows results as a function
of temperature at the ordering wave vector $\bm{Q}$.
Simulations were started in the mean-field configuration to
reduce warm-up times. At low temperatures, the susceptibility grows with
increasing $L$, signaling long-range VBS order. On our largest lattice
size ($L=12$), we observe a sudden drop of the signal at $T\approx 0.06$ (see
Fig.~\ref{fig:Mean-Field}(a)). The energy $\langle \hat{H} \rangle$ shows a
kink at the same temperature [Fig.~\ref{fig:Mean-Field}(b)]. Above this critical
temperature, thermal fluctuations destroy the long-range order.

\subsection{Finite phonon frequencies}

\subsubsection{Equal-time and static quantities}

To map out the phases as a function of phonon frequency, we computed the spin-spin
correlations
\begin{equation}
  S_{S}  (\bm{q},\tau) =   \left<  \hat{S}_{z}(\bm{q},\tau)   \hat{S}_{z}(- \bm{q}) \right>   -  \left<  \hat{S}_{z}(\bm{q},\tau) \right> \left<  \hat{S}_{z}(- \bm{q}) \right>,
\end{equation}
as well as the imaginary-time-displaced correlations of the bond-kinetic energy defined in Eq.~(\ref{eq:Kin_corr}). Here,
\begin{equation}
  \hat{S}_{z}(\bm{q})   = \frac{1}{\sqrt{N}} \sum_{\bm{i}} e^{i \bm{q}\cdot \bm{i}}  \left(  \hat{n}_{\bm{i},\uparrow} -   \hat{n}_{\bm{i},\downarrow}  \right). 
\end{equation}
We also considered the bond-kinetic susceptibility of Eq.~(\ref{eq:Kin_suscep})
and the equivalent form of the spin susceptibility,
$\chi_{S}(\bm{q})$. Because of the O(4) symmetry of Eq.~(\ref{eq:SSH}), see
Sec.~\ref{Symmetries.sec}, the three components of the spin-spin correlations are
degenerate with CDW and s-wave SC correlations. Here, we will discuss the
results from the point of view of spin-spin correlations.

\begin{figure}[b]
  \includegraphics[width=0.49\linewidth]{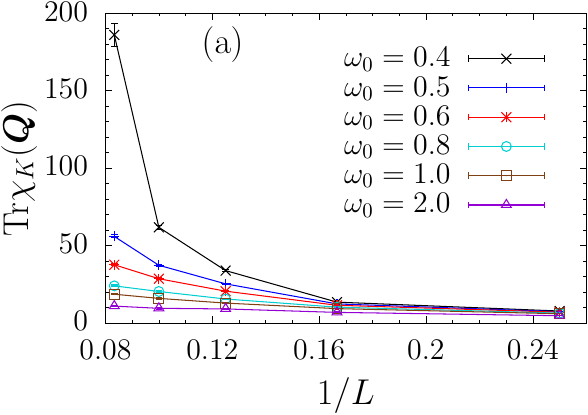} \\
  \includegraphics[width=0.49\linewidth]{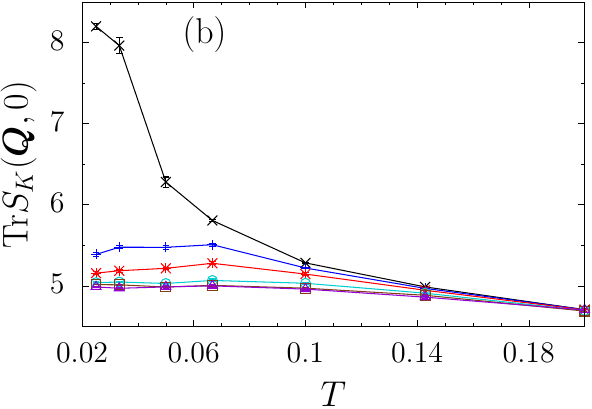}
  \includegraphics[width=0.49\linewidth]{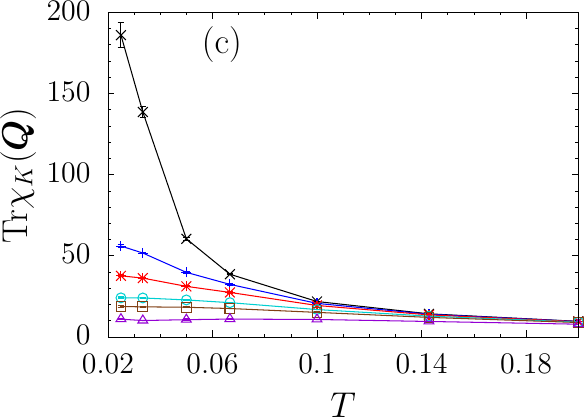}
  \caption{\label{fig:Kin_eq_chi} (a) Finite-size scaling of the bond-kinetic
    susceptibility at $T =1/40$. (b) Temperature dependence of the
    bond-kinetic structure factor and (c) of the bond-kinetic susceptibility for $L=12$.}
\end{figure}

\begin{figure}[h]
  \includegraphics[width=0.49\linewidth]{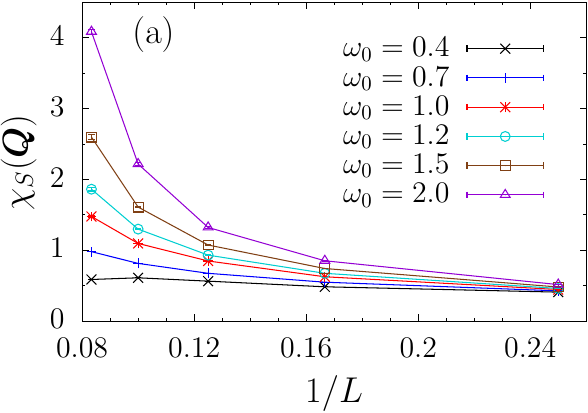} \\
  \includegraphics[width=0.49\linewidth]{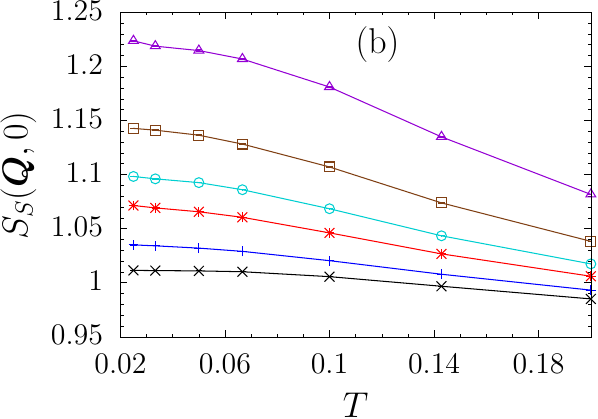}
  \includegraphics[width=0.49\linewidth]{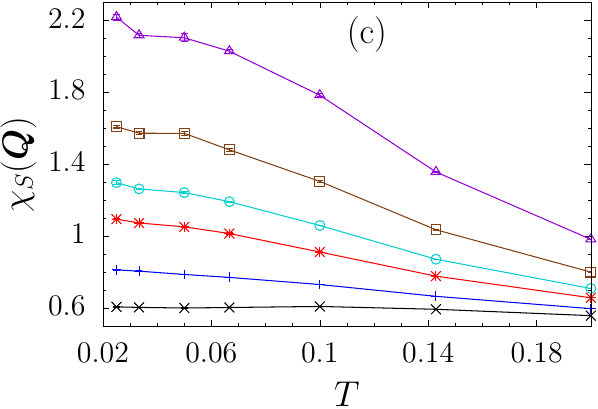}
  \caption{\label{fig:Spin_eq_chi} (a) Finite-size scaling of the spin susceptibility
    at $T= 1/40$. (b) Temperature dependence of the spin structure
    factor for $L=10$. (c) Temperature
    dependence of the spin susceptibility for $L=10$.}
\end{figure}
 
In Fig.~\ref{fig:Kin_eq_chi}, we present the dependence of the bond-kinetic susceptibility and structure factor
on lattice size, temperature, and phonon frequency.  At the
lowest frequency considered ($\omega_0=0.4$), $\text{Tr
}\chi_K(\bm{Q})$ grows as  a function of size and
inverse temperature, suggesting the same $(\pi,\pi)$ VBS order as in
the adiabatic limit. Note that the lowest temperature, $T={1}/{40}$, was not
sufficient to achieve convergence of
$\text{Tr }\chi_K(\bm{Q} )$ for $L=12$.
Contrary to theoretical expectations based on the
log divergence caused by Fermi surface nesting, a nonzero critical value for VBS
order was reported in Ref.~\cite{Xing21} for $\omega_0=1$. While the limitations
regarding system size do not allow us to address this contradiction, 
our observation of VBS order at the dimensional coupling $\lambda={g^2}/{8 k t} = 0.141$ and
for $\omega_0=0.4$ is compatible with VBS order for
$\lambda\gtrsim 0.112$ and $\omega_0=1$ in Ref.~\cite{Xing21}.
As $\omega_0$ is increased, we observe a rapid drop in
$\text{Tr} \chi_K (\bm{Q} )$ that indicates that the
VBS state gives way to another phase.

\begin{figure}[b]
  \includegraphics[width=0.6\linewidth]{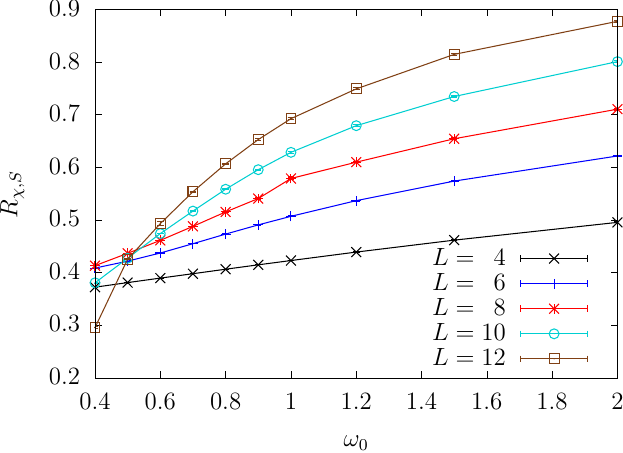}
  \caption{\label{fig:Spin_R_chi} Correlation ratio based on the
    spin susceptibility as defined in Eq.~(\ref{Corr_ratio1.eq}).  Here, $T = 1/40$, which is representative of
    the ground state for this quantity.}
\end{figure}

In Fig.~\ref{fig:Spin_eq_chi}(a),   we  show results for the spin degrees of
freedom. At low temperatures,  the size-dependence of the AFM
spin susceptibility shows a marked increase at \textit{high} phonon
frequencies. In Figs.~\ref{fig:Spin_eq_chi}(b) and (c),  the temperature
dependence at fixed lattice size shows that we are able to achieve convergence
with respect to temperature. This allows us to compute  the correlation ratio
\begin{equation}\label{Corr_ratio1.eq}	
R_{\chi,S}    =  1 - \frac{\chi_{S}(\bm{Q} + \Delta \bm{q} ) } { \chi_{S}(\bm{Q}  ) }
\end{equation}
where  $ |\Delta \bm{q}  | = 2 \pi/L $. This renormalization group invariant
quantity takes the value of unity (zero) in the ordered (disordered) phase.
At $T=0$ and for a continuous transition, it scales as
\begin{equation}
\label{Corr_ratio.eq}	
  R_{\chi,S}     =  f \left(   [\omega_0 - \omega_0^c] L^{1/\nu} \right). 
\end{equation}

Figure~\ref{fig:Spin_R_chi} shows $R_{\chi,S}$ as a function of system size
for the lowest temperature available (representative  of the ground state).
Although corrections to scaling,  not included in Eq.~(\ref{Corr_ratio.eq}),    lead
to a drift of the crossing points, the data suggest a critical phonon
frequency $\omega_0^c \simeq 0.6$  for the onset of long-range AFM order.

\begin{figure}[t]
  \includegraphics[width=0.6\linewidth]{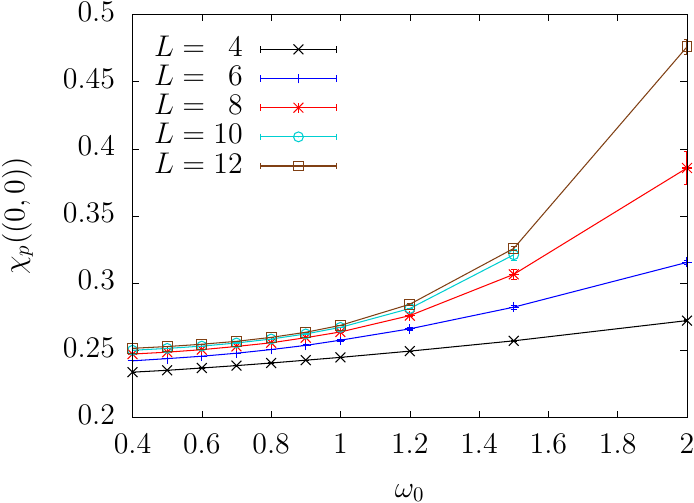}
  \caption{\label{fig:Parity} Parity susceptibility [Eq.~(\ref{Parity_suscep.eq})] at $T = 0.1$.}
\end{figure}
 
Being a modulation of the bond-kinetic energy, the VBS state does
not break the underlying O(4) symmetry of the lattice.  However, it does break
translation and rotation symmetries. On the other hand, the AFM phase does break
the O(4) symmetry, as can be demonstrated by  computing the
susceptibility of the parity operator defined in Eq.~(\ref{eq:par}),
\begin{equation}
  \label{Parity_suscep.eq}
  \chi_p(\bm{q})   =   \int_{0}^{\beta} \text{d}\tau \sum_{\bm{r}} e^{i \bm{q} \cdot \bm{r} } \langle  \hat{p}_{\bm{r}}(\tau)  \hat{p}_{\bm{0}}  \rangle.
\end{equation}
Since $\hat{p}_{\bm{i}}$ is an Ising variable that changes sign under an
$ O(4)$ transformation $\mathcal{M}$ with $\det{\mathcal{M}} = -1$, we expect
$ \chi_p( \bm{0}) $ to diverge at a critical temperature associated with
a phase transition in the 2D Ising universality class.  Being an 8-point correlation
function, $\chi_p(\bm{q}) $ becomes very noisy at low temperatures and we are
restricted to $T= 0.1$. Figure~\ref{fig:Parity} shows results as 
a function of system size and phonon frequency.  For
$\omega_0 =2$, $\chi_p(\bm{0}) $ grows with increasing $L$,
suggesting that for this frequency the Ising temperature is below
$T = 0.1$. On the other hand, for $\omega_0 = 1$ (still in the AFM phase) our
temperature is too high to capture the Ising transition. We conclude that the
AFM phase breaks the O(4) symmetry down to SO(4) at a finite-temperature Ising
transition occurring at $T_c^{I}$.  A natural conjecture is that $T_c^{I}$
vanishes at $\omega_0^{c}$.
  
\subsubsection{Dynamical quantities}

\begin{figure}[t]
  \includegraphics[width=1\linewidth]{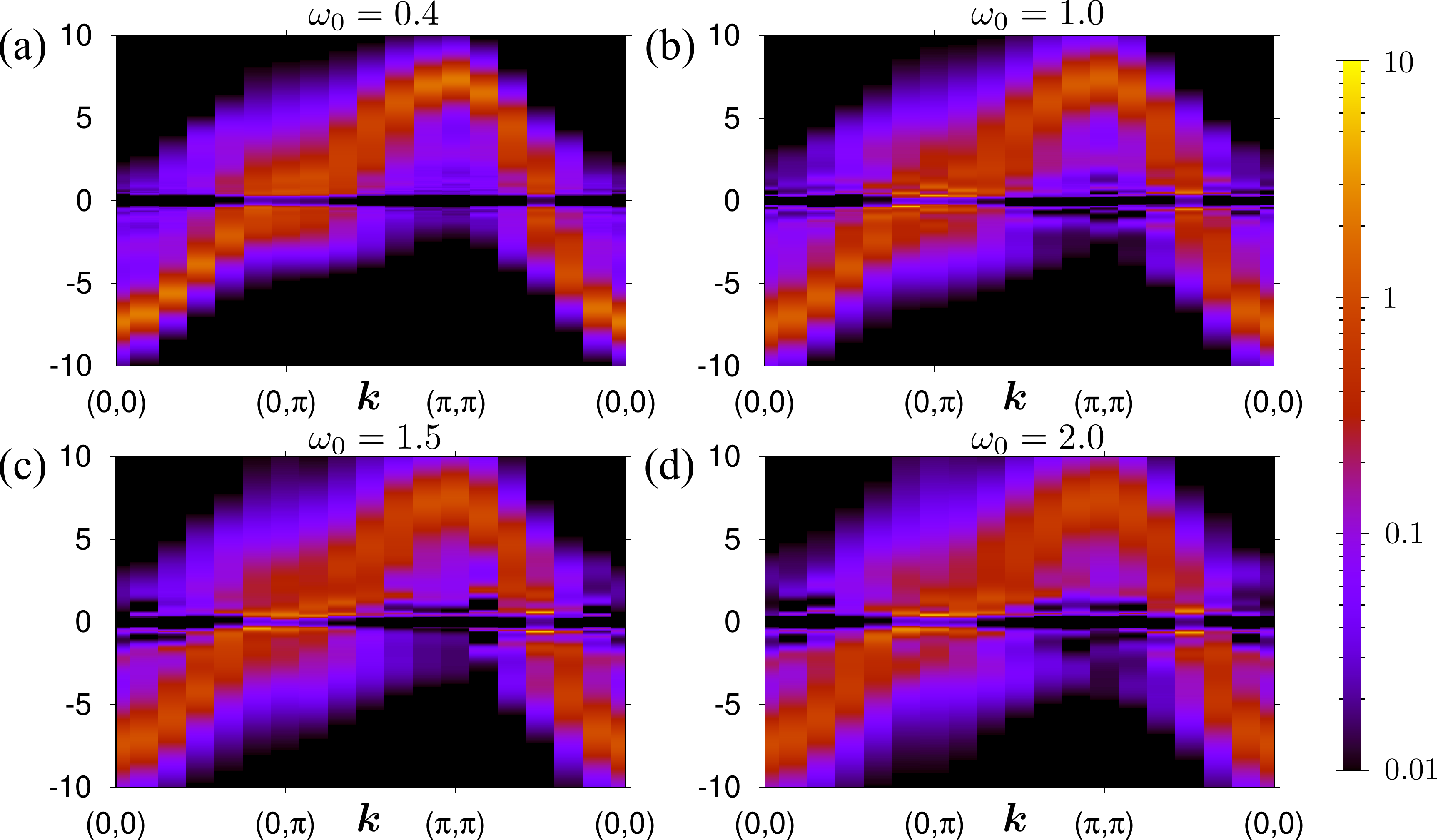}
  \caption{ \label{Akom.fig} Single-particle spectral function $A(\bm{k},\omega)$ for different
    phonon frequencies. Here, $L=12$, $\beta = 40$.}
\end{figure}

To extract spectral functions from QMC data for imaginary-time correlators via analytic
continuation, we used the ALF  
implementation \cite{ALF_v2} of the stochastic maximum entropy algorithm
\cite{Sandvik98,Beach04a}.

The single-particle spectral function $A(\bm{k},\omega) $, accessible in ARPES
experiments, is related to the imaginary-time Green function via
\begin{equation}
  \langle \hat{c}^{\phantom\dagger}_{\bm{k}.\sigma} (\tau)
  \hat{c}^{\dagger}_{\bm{k},\sigma} (0)   \rangle   =  \frac{1}{\pi} \int
  \text{d} \omega  \, \frac{e^{-\tau \omega} }  {  1 + e^{-\beta\omega} }
  A(\bm{k}, \omega).
\end{equation}

Figure~\ref{Akom.fig}(a) shows $A(\bm{k},\omega)$ for $\beta = 40$ and
$L=12$. The coupling of the Einstein phonon mode to the
electrons breaks the $\hat{Q}_b \rightarrow -\hat{Q}_b$ symmetry. Consequently,
$\frac{1}{2N} \sum_{b}\big< \hat{Q}_b \big> $ acquires a non-zero expectation
value that renormalizes the electronic bandwidth.  For $ \omega_0 = 0.4$,
$\frac{1}{2N} \sum_{b}\big< \hat{Q}_b \big> = -0.56510(7) $, yielding an 
effective hopping $t_{\text{eff}} = 1.85$.
This explains the observed range of the band from $-4 t_{\text{eff}} $ at
$\bm{k}=(0,0)$ to $4 t_{\text{eff}} $ at $\bm{k}=(\pi,\pi)$.  For $ \omega_0 =
0.4$, inside the VBS phase, the $(\pi,\pi)$ modulation of the hopping opens a gap
at the non-interacting Fermi surface, as visible for $\bm{k} =(0,\pi) $
and $\bm{k} =(\pi/2,\pi/2) $ in Fig.~\ref{Akom.fig}(a).  Both, the gap and
the cosine band of width $8t_{\text{eff}}$ are features that can be
qualitatively accounted for at the mean-field level.  However, the spectral
function exhibits low-lying spectral weight that extends over the considered
path in the Brillouin zone. In analogy with the 1D Holstein model
\cite{Assaad08}, and guided by the results of a self-consistent Born
  approximation shown below, we
attribute this low-energy feature to polaron formation. The O(4) symmetry of the model implies 
$A(\bm{k},\omega) = A(\bm{k}+\bm{Q},-\omega)$.
Hence, in the absence of symmetry breaking, the polaron band is nested and
should exhibit instabilities to AFM or VBS order.

Figure~\ref{SCBA.fig} shows the single-particle 
spectral function from a self-consistent Born approximation (see appendix).
In the latter, we neglect phonon renormalization and instead use an
effective hopping $t=1.85$ derived from the QMC data.
Figure~\ref{SCBA.fig}(a) reveals a cosine band with a gap of the order of the
bare phonon frequency $\omega_0$ that is crossed by a narrow polaron band. A full
gap---as in the QMC data---is achieved within this approximation
by an additional single-particle term
\begin{equation}\label{eq:symm_break}
\hat{H}_{\lambda}= \lambda \sum_{\i}  \text{e}^{\text{i}\boldsymbol{Q} \i} \left(\hat{n}_{\i,\uparrow}-\hat{n}_{\i,\downarrow} \right)
\end{equation}
that explicitly breaks the O(4) symmetry by enforcing AFM order, see Fig.~\ref{SCBA.fig}(b).
The Born approximation
  provides a qualitative interpretation of the features observed numerically
  but---as expected---does not capture quantitative aspects such as the true size of
  the gaps.

\begin{figure}[t]
  \includegraphics[width=1\linewidth]{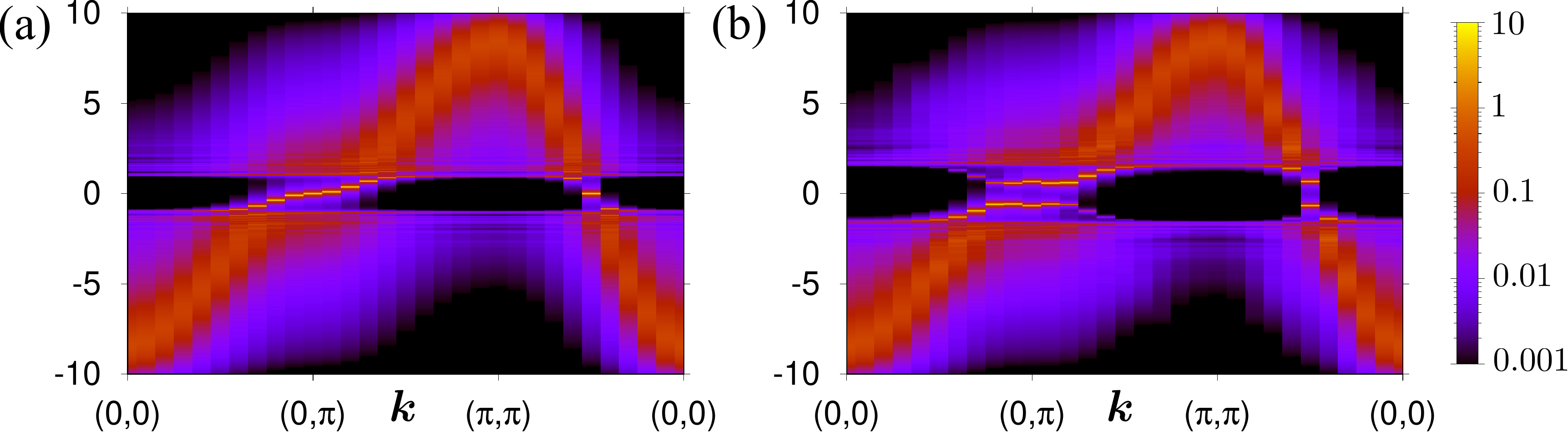}
  \caption{ \label{SCBA.fig} (a) Single-particle spectral function $A(\bm{k},\omega)$
  from a self-consistent Born approximation for
  $\beta=40$, $t=1.85$, $g=1.5$, $\omega_0=1.0$. (b) As in (a) but with broken O(4)
  symmetry via addition of $\hat{H}_\lambda$ [Eq.~(\ref{eq:symm_break})] with $\lambda=0.5$.}
\end{figure}

Upon increasing the phonon frequency to $\omega_0=2$,
Figs.~\ref{Akom.fig}(b)-(d), $\frac{1}{2N} \sum_{b} \big< \hat{Q}_b \big>$
remains almost constant and, consequently, the width of the cosine band does
not change substantially.  However, we observe a transfer of spectral
weight to the polaron band.  The origin of the gap at $ \omega_0 > \omega_0^{c} $ is to
be found in AFM ordering. Our QMC data suggest that the single-particle gap
remains open across the VBS-AFM transition. 

\begin{figure}[t]
  \includegraphics[width=\linewidth]{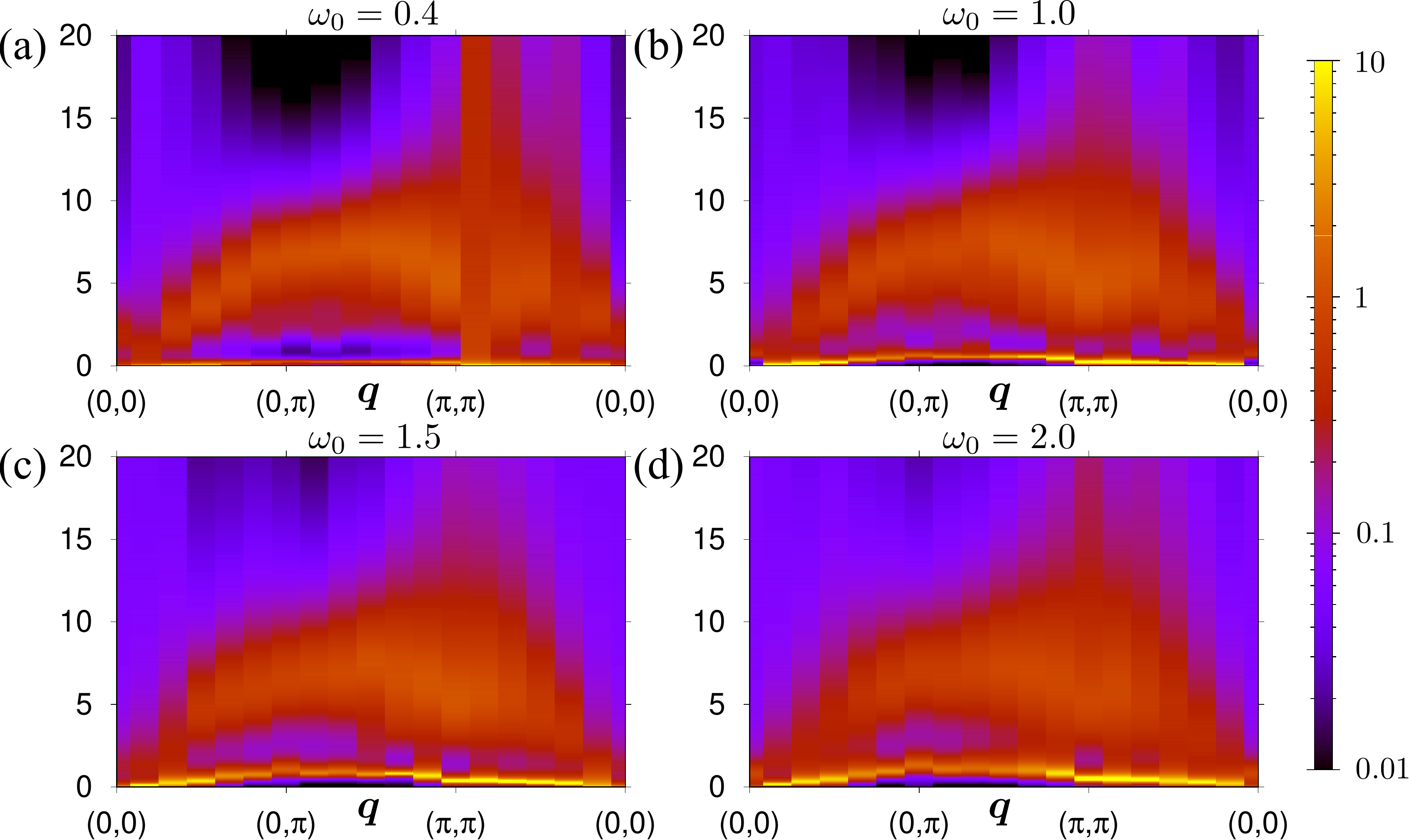}
  \caption{ \label{Kom.fig} VBS dynamical structure factor $S_K(\bm{q},\omega)$ for different phonon frequencies.
     Here, $L=12$, $\beta = 40$. }
\end{figure}

\begin{figure}[b]
  \includegraphics[width=\linewidth]{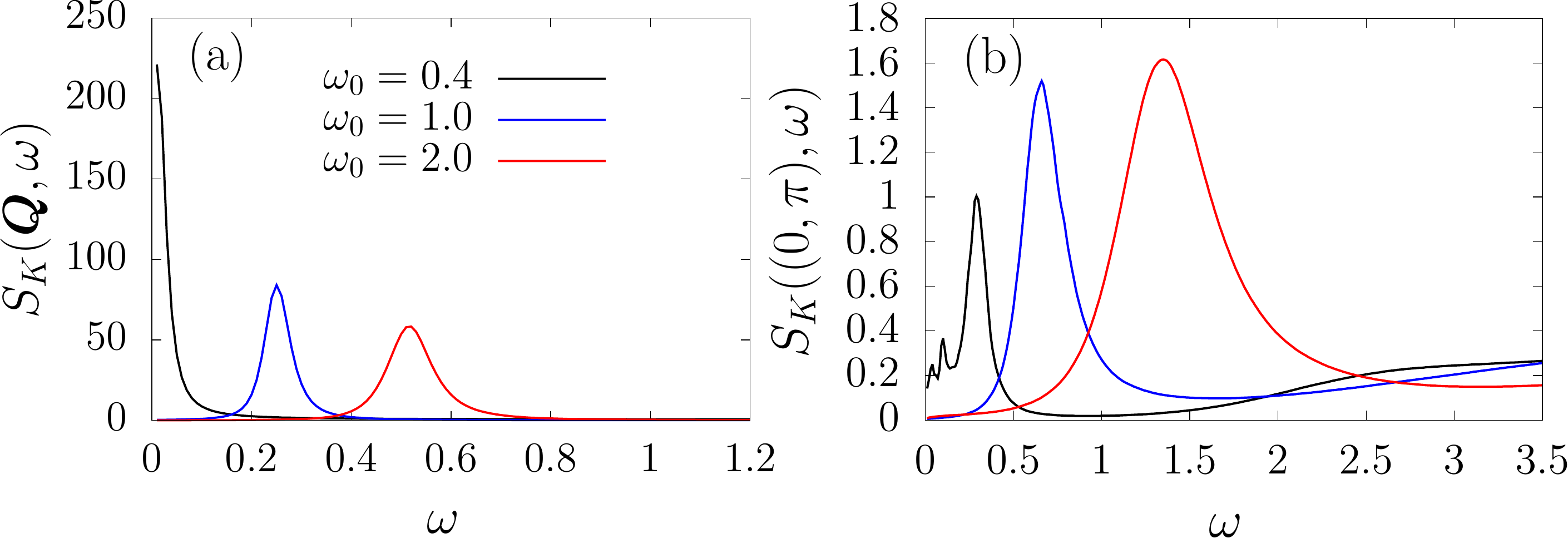}
  \caption{ \label{Spectral_Kin.fig} VBS dynamical structure factor for (a)
    $\bm{q}=\bm{Q}$ and  (b) $\bm{q}=(0,\pi)$ for different $\omega_0$.  Here, $L=12$, $\beta = 40$.}
\end{figure}

Figure~\ref{Kom.fig} shows the VBS dynamical structure factor
$S_K(\bm{q},\omega) $ at four different phonon frequencies.  We computed the
imaginary part of the dynamical VBS susceptibility $\text{Tr}
\chi_K''(q,\omega)$ by using the maximum entropy method \cite{ALF_v2} to invert
\begin{equation}
  \text{Tr} S_K (q,\tau) = \frac{1}{\pi}
  \int \text{d} \omega \, \frac{e^{- \tau \omega} }{ 1 - e^{-\beta  \omega} } \, \text{Tr} \chi_K''(q,\omega). 
\end{equation}
The dynamical VBS structure factor then follows from
\begin{equation}
  S_K (\bm{q},\omega) = \frac{\text{Tr} \chi_K''(q,\omega)} {\left( 1 - e^{-\beta  \omega} \right) }\,. 
\end{equation}
Since the phonons couple to the bond-kinetic energy, 
$S_K (\bm{q},\omega)$ should reveal both the phonon dynamics and the
particle-hole continuum.  At $\omega_0 = 0.4$ (Fig.~\ref{Kom.fig}(a)), we see
substantial, very low-lying weight as well as high-energy features that reflect the
particle-hole continuum.  The low-lying excitation corresponds to the phonon
mode. The fact that it is soft at $\bm{q}=\bm{Q}$ is a signature of long-range VBS order.
In Fig.~\ref{Spectral_Kin.fig} we show $S_K (\bm{q},\omega)$ at wave vector 
$\bm{Q}$ and also at $\bm{q}=(0,\pi)$ for comparison.
However, we cannot resolve the dispersion relation.  In comparison to
the bandwidth, the renormalized phonon modes are very slow and are at the
origin of long autocorrelation times. In the AFM phase,
Figs.~\ref{Kom.fig}(c)-(d), the phonon mode acquires a gap.

\begin{figure}[t]
  \includegraphics[width=\linewidth]{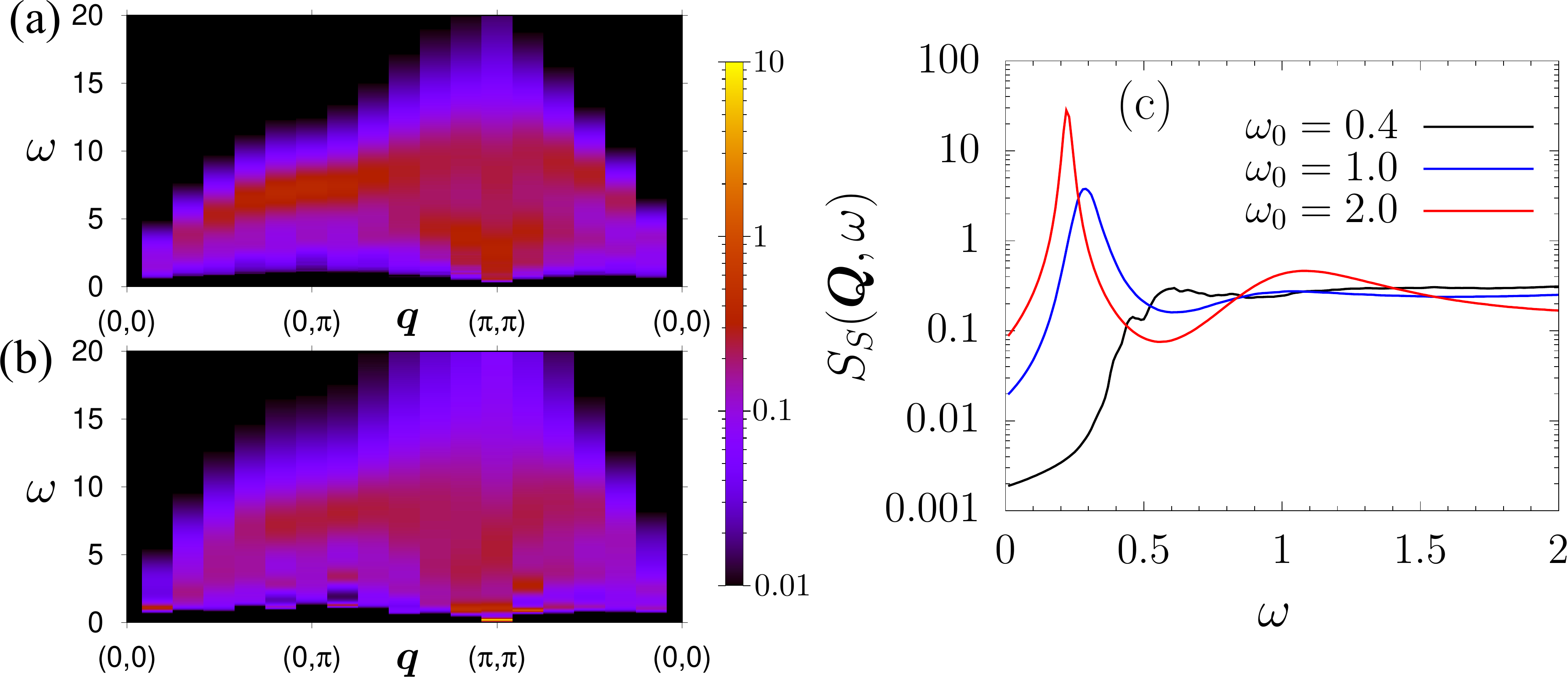}
  \caption{ \label{Som.fig} Spin dynamical structure factor
    $S_S(\bm{q},\omega)$ for (a)
    $\omega_0=0.4$, (b) $\omega_0 =2.0 $, (c) fixed $\bm{q}=\bm{Q}$ and different $\omega_0$.  Here, $L=12$,  $\beta = 40$. }
\end{figure}

Finally, we show the dynamical spin structure factor
\begin{equation}
  S_S (\bm{q},\omega) = \frac{\text{Tr} \chi_S''(q,\omega)} {\left( 1 - e^{-\beta  \omega} \right) }\,. 
\end{equation}
in Fig.~\ref{Som.fig}. Because phonons do not carry spin, they are not
visible in spin-flip scattering processes. According to Fig.~\ref{Som.fig}(a),
in the VBS phase, $S_S(\bm{q},\omega)$ is dominated by the
particle-hole continuum.  For $\omega_0\geq 1.0$, low-energy spectral weight
at $\bm{q}=\bm{Q}$ reflects long-range AFM order, see Figs.~\ref{Som.fig}(b)-(c).

\section{Discussion and conclusions}\label{Summary.sec}

Our results provide both algorithmic and physical insights into the fundamental
2D SSH model.

\subsection{Langevin dynamics}

We used a Langevin dynamics updating scheme with Fourier acceleration
\cite{Batrouni19} in the framework of the auxiliary-field quantum Monte Carlo method.

In contrast to the HMC approach in Ref.~\cite{Beyl17}, we computed forces
exactly for a given field configuration.  A comparison between stochastic and
deterministic calculations of forces can be found in Ref.~\cite{Ulybyshev19b}.
Although the CPU time per sweep is longer and scales as $L^6 \beta$,
fluctuations, especially for time-displaced correlation functions, are smaller.

One of the key difficulties encountered in Langevin dynamics are zeros of the
determinant, which lead to logarithmic singularities of the action.  Exploiting the
O(4) symmetry of the SSH model, the determinant can be written as the square of a
Pfaffian, whose average sign provides a measure for the density
of zeros of the determinant.  We demonstrated that for \textit{low} phonon frequencies, the
density is small, so that Langevin simulations can be stabilized
using an adaptive time step scheme. Nevertheless, simulations occasionally suffer
from spikes in observables when the stochastic walk approaches a zero.
Obviously, such configurations have very small weight and a hybrid
molecular dynamics update may be more efficient. In principle,
the resulting ergodicity issues can be overcome
by a \textit{complexification} of the fields \cite{Beyl17}.  For the Hubbard
model, this is possible since the decoupling of the interaction can be done
in various channels. For the SSH model, we do not have such liberty.

We find that global Langevin updates are a good choice in the
adiabatic regime where local moves fail. As the phonon frequency grows,
Langevin dynamics becomes increasingly challenging. At the same time, local
updates become favorable, as discussed, \eg, in Ref.~\cite{Hohenadler14}.

\subsection{Physics of the 2D SSH model}

Despite its apparent simplicity and fundamental
nature, remarkably little was known about the 2D SSH model. The existence and
type of long-range VBS order was settled only recently \cite{Xing21}.
Our results elucidate the physics of the SSH model at a fixed
electron-phonon coupling  and as a function of the phonon frequency. 
In addition to static observables, we specifically also presented excitation
spectra from QMC and analytical methods.  The numerical results were further
complemented with a mean-field approach to the VBS phase and simulations showing the
temperature-driven destruction of VBS order in the adiabatic limit.

The features of the phase diagram in Fig.~\ref{Phase_diagram.fig} are tied to the O(4) symmetry
of the model.  In particular, the single-particle spectral function satisfies
$A(\bm{k},\omega) = A(\bm{k} + \bm{Q},-\omega)$.  Hence, any Fermi liquid state that does not break
this symmetry will ultimately be unstable to orders that can open up a gap.
This includes the $\bm{Q}=(\pi,\pi)$ VBS phase as well as AFM order.  In
the adiabatic limit, the problem simplifies since the phonons become
classical and mean-field theory gives a $(\pi,\pi)$ VBS phase as the exact
ground state. For $\omega_0>0$, the phonons can be integrated out in favor of a
retarded interaction. The latter reduces to Eq.~(\ref{Kb2.eq}) in the
antiadiabatic limit $\omega_0\to\infty$, where it triggers an AFM state that is
degenerate with CDW and SC states. Two key results are the existence of AFM order
down to the experimentally relevant adiabatic regime $\omega_0< t$ and a 
direct transition from VBS to AFM order. The single-particle spectral function
supports the picture of a narrow polaronic band undergoing
a transition from a $(\pi,\pi)$ VBS to AFM/CDW/SC.  In the particle-hole
channel, the dynamical VBS correlation function reveals the phonon dynamics
as a function of decreasing phonon frequency, including a softening at $(\pi,
\pi)$.  On the other hand, as $\omega_0$ grows, we observe enhanced spectral
weight at low energies and at $\bm{q}=\bm{Q}$ in the spin channel.

\begin{figure}[t]
\includegraphics[width=0.5\linewidth]{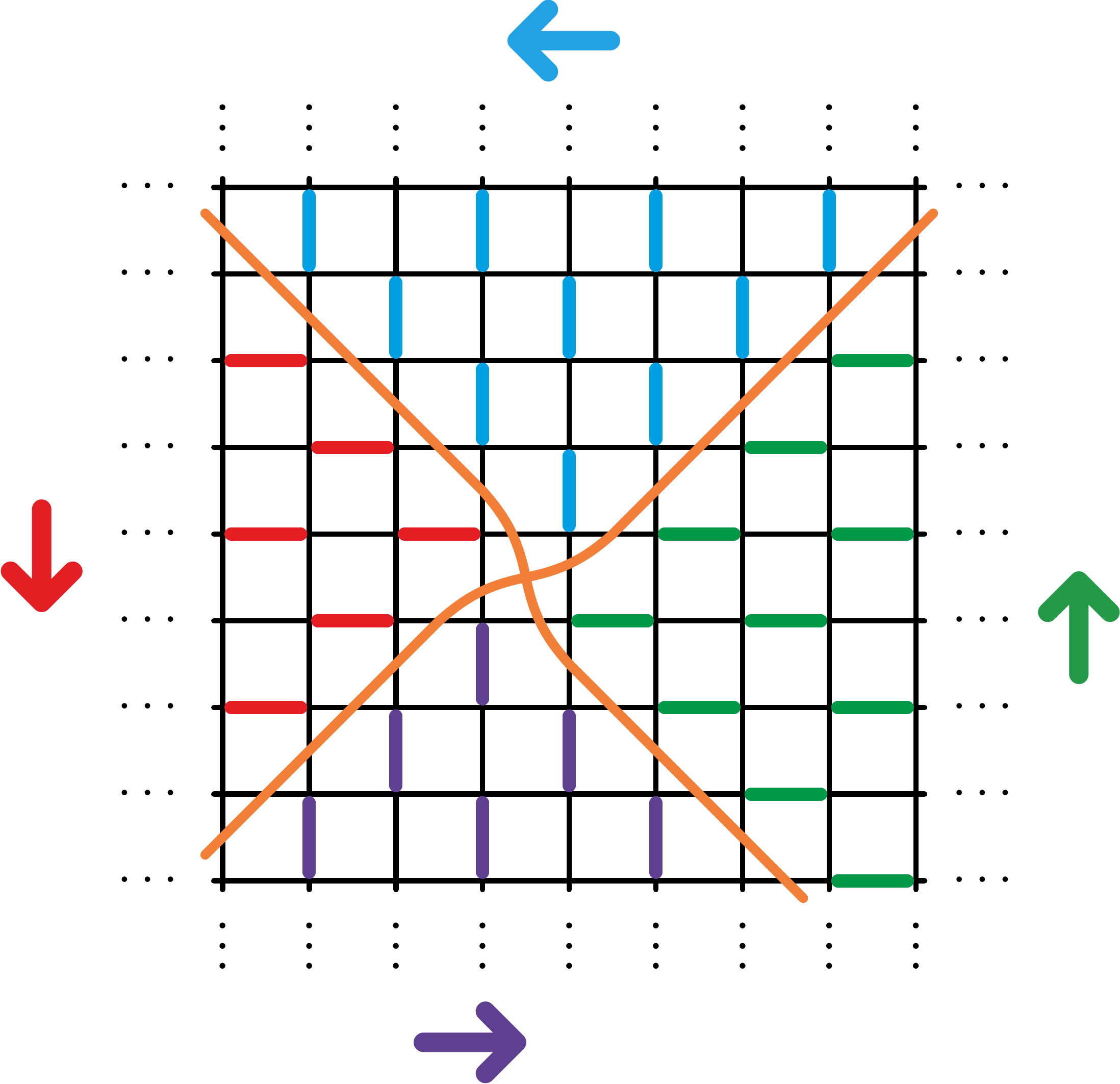}
\caption{ \label{C4_vortex.fig} A vortex of the $(\pi,\pi)$ VBS state.
  Arrows represent the four degenerate VBS patterns that break the C$_4$ symmetry.
  By crossing a domain wall (orange lines), the angle changes by $\pi/2$. A full
  circle around the core yields $2\pi$. This {\em trivial} vortex does
  not carry a spin-1/2 degree of freedom, in contrast to the case of $(0,\pi)$ or
   $(\pi,0)$ VBS orders \cite{Levin04}.}
\end{figure}

The VBS phase breaks lattice symmetries but not the above O(4) symmetry.
On the other hand, AFM or SC/CDW phases break the O(4) symmetry down to SU(2)
but leave lattice symmetries in tact. Starting at high temperatures, the
symmetry reduction to SU(2) occurs in two steps. At a critical
temperature, spontaneous ordering of the parity operator takes place at an
Ising transition. Even parity corresponds to the SC/CDW phase, odd parity to the
AFM phase. Then, at $T=0$, the SU(2) spin (pseudospin) symmetry is spontaneously
broken, leaving the SU(2) pseudospin (spin) symmetry unbroken.

Generically, the O(4) symmetry will be broken down to SU(2) by,  for
example, adding a next-nearest-neighbor hopping.   In this case, we
expect the phase diagram to be dominated by superconductivity, as
the Cooper instability is insensitive to
the shape of  the Fermi surface.  The stability of the VBS phase as a
function of an O(4) symmetry-breaking interaction  such as a chemical potential
or a next-nearest-neighbor hopping deserves a detailed investigation. 

The nature of the VBS-AFM transition remains elusive. 
Because it occurs between states with different broken symmetries,
  Ginzburg-Landau  order  parameter  theory   generically predicts either a  first-order
transition or a region of coexistence. Within the $\omega_0$ resolution of our
results, this was not observed. Instead, the transition appears continuous. The theory of deconfined
quantum critical points (DQCPs) \cite{Senthil04_1,Senthil04_2} does not apply.
To see this, we can adopt the DQCP picture of an 8-component Dirac metal with
five anti-commuting  AFM and  VBS mass terms  \cite{Tanaka05,Senthil06,Liu18}.
The algebra of the mass terms guarantees that the core of a vortex in the VBS
order parameter carries a spin-1/2  excitation. However, this requires an
ordering wavevector $(0,\pi)$ or $(\pi,0)$. In contrast,
the $(\pi,\pi)$ VBS observed here does not correspond to a Dirac mass term.
This point of view is  substantiated by  noticing that a $C_4$  vortex of the
$(\pi,\pi)$ VBS can be trivial, as explained in Fig.~\ref{C4_vortex.fig}.
Finally, a deconfined VBS-CDW phase transition as a function of phonon
frequency exists in the spinless 1D SSH model \cite{Weber19}, whereas long-range
AFM order is ruled out in 1D models by the Mermin-Wagner theorem.

In summary, we have established the existence of a  $(\pi,\pi)$-ordered VBS
phase and an AFM phase in the 2D SSH model with quantum phonons by means of QMC simulations. Notably, the
AFM phase exists even at finite phonon frequencies. We observed an apparently
direct transition between these phases with no signatures of an intermediate
metallic region. Finally, we provided an interpretation of the numerical results
for the single-particle spectral function in terms of gap formation in a narrow
polaronic band.

{\em Note added:}
During the preparation of this article we became aware of
Ref.~\cite{Cai21}, the results of which appear to be fully consistent with ours.
While the authors do not present excitation spectra, they provide a
phase diagram with critical values for multiple parameter sets based on larger
lattice sizes than the present work. Nevertheless, our critical value
$\omega_0^c\approx 0.6$ for the dimensionless coupling constant
$\lambda={g^2}/{8 k t}\approx 0.141$ is in satisfactory agreement with their
phase boundary. Moreover, Ref.~\cite{Cai21} also points out that the critical
coupling for VBS order is expected to vanish in the 2D SSH model, in contrast to
the findings of Ref.~\cite{Xing21}. Finally, the authors of Ref.~\cite{Cai21}
provide similar arguments regarding the properties of vortices of the VBS
pattern and its implications for the interpretation of the VBS-AFM transition.

\begin{acknowledgments}
We thank E. Huffman for helpful discussions on the
calculation of the Pfaffian and F. Goth for discussions on related work.  The
authors gratefully acknowledge the Gauss Centre for Supercomputing
e.V. (www.gauss-centre.eu) for funding this project by providing computing
time on the GCS Supercomputer SUPERMUC-NG at Leibniz Supercomputing Centre
(www.lrz.de).  FFA thanks the W\"urzburg-Dresden Cluster of Excellence on
Complexity and Topology in Quantum Matter ct.qmat (EXC 2147, project-id
390858490), AG and SB the DFG funded SFB 1170
on Topological and Correlated Electronics at Surfaces and Interfaces.
\end{acknowledgments}

\appendix*
\section{Self-consistent Born approximation}\label{app:B}

The Dyson equation for the full Green function is
\begin{equation}\label{eq:dyson}
G(\bm{k},\text{i}\omega_m) = \left[G_0^{-1}(\bm{k},\text{i}\omega_m) - \Sigma(\bm{k},\text{i}\omega_m)\right]^{-1}
\end{equation}
with the non-interacting Green function $G_0$ and fermionic Matsubara frequencies $\omega_m$.
Here, we only consider the Fock contribution to the electron self energy,
\[ 
~\Sigma=~
\vcenter{\hbox{\begin{tikzpicture}
  \begin{feynman}
    \vertex (i);
    \vertex[right=1.5cm of i] (o);
    \diagram*{
     (i) -- [double,thick,with arrow=0.5,arrow size=0.2em]  (o) --  [ boson, half right] (i)
         };
  \end{feynman}
\end{tikzpicture}}}~+\dots,~
\]
where the wavy line represents the non-interacting phonon propagator $D_0$ and the double line the full Green function $G$.
The contribution of this Feynman diagram is given by \cite{Jishi13,Heid17}
\begin{eqnarray}\label{eq:sigma}
&&\Sigma (\bm{k},\text{i} \omega_m) = \\
&& -\frac{1}{\beta} \sum_{\text{i}\omega_n} \sum_{ \bm{q}, \bm{\delta}}
\left| g^{\bm{q},\bm{\delta}}_{\bm{k}+\bm{q},\bm{k}}\right|^2  G(\bm{k}+\bm{q},\text{i}\omega_n) D_0(\text{i}\omega_n-\text{i} \omega_m)\,, \nonumber
\end{eqnarray}
to be solved self-consistently together with Eq.~(\ref{eq:dyson}). 

The matrix elements $g^{\bm{q},\bm{\delta}}_{\bm{k}+\bm{q},\bm{k}}$ are
defined by the electron-phonon interaction in Hamiltonian~(\ref{eq:SSH}), which can
be expressed after Fourier transformation as
\begin{eqnarray}
\hat{H}_{\text{ep}} &=&  \sum_{\bm{k},\bm{q},\sigma} \sum_{\boldsymbol{\delta}} g^{\bm{q},\bm{\delta}}_{\bm{k}+\bm{q},\bm{k}}
 \hat{c}_{\bm{k}+\bm{q},\sigma}^{\dagger} \hat{c}_{\bm{k},\sigma}^{\phantom{\dagger}}  \left(\hat{d}_{-\bm{q},\boldsymbol{\delta}}^{\dagger} +\hat{d}_{\bm{q},\boldsymbol{\delta}}^{\phantom{\dagger}}  \right), \nonumber \\
g^{\bm{q},\bm{\delta}}_{\bm{k}+\bm{q},\bm{k}} &=&
\frac{g}{\sqrt{2 m \omega_0 N}} 
  \left(\text{e}^{-\text{i} (\bm{k}+\bm{q}/2) \boldsymbol{\delta}} 
 + \text{e}^{\text{i} (\bm{k}+\bm{q}/2) \boldsymbol{\delta}}  \right)\, .
\end{eqnarray}
Here, we have rewritten the position operator of the phonons in terms of
bosonic creation and annihilation operators, $\hat{Q}_b = \frac{1}{\sqrt{2 m
    \omega_0}} \left( \hat{d}_{b}^{\dagger} + \hat{d}_b^{\phantom{\dagger}}
\right)$. The vector $\boldsymbol{\delta}$ connects two nearest-neighbor
sites, $b=\langle \i,\i+\boldsymbol{\delta}\rangle$. The non-interacting
phonon propagator can be written as
\begin{equation}
D_0(\text{i}\Omega_m) = \frac{1}{\text{i}\Omega_m - \omega_0} - \frac{1}{\text{i}\Omega_m + \omega_0}\,,
\end{equation}
where $\Omega_m$ is a bosonic Matsubara frequency. To carry out the summation
over the Matsubara frequencies in Eq.~(\ref{eq:sigma}), we rewrite the Green
function with the spectral function $A(\bm{k},\omega)=-({1}/{\pi})\text{Im}\,
G^{\text{R}}(\bm{k},\omega)$,
\begin{align}
G(\bm{k},\text{i}\omega_m) = \int \text{d} \omega \frac{A(\bm{k},\omega)}{\text{i}\omega_m-\omega},
\end{align}
where the retarded Green function
$G^{\text{R}}(\bm{k},\omega)=G(\bm{k},\text{i}\omega_m\rightarrow\omega+\text{i}\varepsilon)$
is obtained by analytical continuation with an infinitesimal $\varepsilon$. We
obtain for the self-energy \cite{Assaad08}
 \begin{eqnarray}\nonumber
\Sigma (\bm{k}, \omega+\text{i}\varepsilon) 
&=& \int \text{d}\overline{\omega} \sum_{ \bm{q}, \bm{\delta}}
\left| g^{\bm{q},\bm{\delta}}_{\bm{k}+\bm{q},\bm{k}}\right|^2 A(\bm{k}+\bm{q},\overline{\omega})  \\ \nonumber
& & \quad\times \left[ \frac{n_B(\omega_0) +n_F(\overline{\omega})}{\omega+\text{i}\varepsilon+\omega_0 - \overline{\omega}} \right.
\\
   &&\quad\quad+ \left.\frac{n_B(\omega_0)+1-n_F(\overline{\omega})}{\omega+\text{i}\varepsilon-\omega_0-\overline{\omega}}\right] .
\end{eqnarray}
Here, $n_B(\omega_0)=\frac{1}{e^{\beta\omega_0}-1}$ and $n_F(\omega)=\frac{1}{e^{\beta\omega}+1}$ are the Bose-Einstein and Fermi-Dirac distributions respectively.

Our approximation neglects the renormalization of the phonons due to the electrons and vertex corrections.

%\bibliography{../fassaad}


\end{document}